# $N_2$-$H_2$ capacitively coupled radio-frequency discharges at low pressure. Part II. Modelling results: the relevance of plasma-surface interaction

**Miguel Jiménez-Redondo*[1,2], Audrey Chatain[3,4], Olivier Guaitella[4], Guy Cernogora[3], Nathalie Carrasco[3], Luis Lemos Alves[5], Luis Marques[1]**

[1]Centro de Física das Universidades do Minho e do Porto, Universidade do Minho, 4710-057, Braga, Portugal
[2]Present address: Instituto de Estructura de la Materia, IEM-CSIC, Serrano 123, 28006 Madrid, Spain
[3]LATMOS/IPSL, UVSQ Université Paris-Saclay, Sorbonne Université, CNRS, 78280 Guyancourt, France
[4]LPP, École Polytechnique, Sorbonne Université, université Paris-sud, CNRS, 91128, Palaiseau, France
[5]Instituto de Plasmas e Fusão Nuclear, Instituto Superior Técnico, Univ. Técnica de Lisboa, Lisboa, Portugal

*e-mail: m.jimenez@csic.es

## Abstract

In this work, we present the results of simulations carried out for $N_2$-$H_2$ capacitively coupled radio-frequency discharges, running at low pressure (0.3–0.9 mbar), low power (5–20 W), and for amounts of $H_2$ up to 5%. Simulations are performed using a hybrid code that couples a two-dimensional time-dependent fluid module, describing the dynamics of the charged particles in the discharge, to a zero-dimensional kinetic module, that solves the Boltzmann equation and describes the production and destruction of neutral species. The model accounts for the production of several vibrationally and electronic excited states, and contains a detailed surface chemistry that includes recombination processes and the production of $NH_x$ molecules. The results obtained highlight the relevance of the interactions between plasma and surface, given the role of the secondary electron emission in the electrical parameters of the discharge and the critical importance of the surface production of ammonia to the neutral and ionic chemistry of the discharge.

## 1. Introduction

Low pressure $N_2$/$H_2$ plasmas are of interest in a wide variety of fields, from fusion to planetary science, and even in technological-driven research their study often involves a fundamental component, to deepen the knowledge on these systems.

In fusion research, the focus lies on the use of $N_2$ to attenuate the local heat loads on tungsten divertors [1]. $N_2$ acts as an efficient extrinsic radiator, but in turn it creates a series of setbacks to the plasma operation, namely the implantation of nitrogen and nitriding of plasma-facing materials, sputtering caused by energetic nitrogen ions, and the formation of tritiated ammonia which conditions the operation of gas plants and cryopumps, and influences the nature of the walls [2, 3]. The study of $NH_3$ formation and of strategies for its mitigation is one of the current priorities in fusion research [4, 5]. If a carbon material is used at the divertor plates, sputtering can lead to the formation of tritiated hydrogenated amorphous carbon in the surfaces [6, 7]. Laboratory experiments in conditions similar to those of the divertor region have shown a reduction in the formation of these compounds when introducing $N_2$ in an $H_2$/$CH_4$ plasma mixture [8].

Plasmas containing $N_2$/$H_2$ are extensively used in technological applications up to the nanoscale level [9], including deposition of thin films [10-12], etching of organic low permittivity films [13, 14], surface treatment by nitriding of metals [15, 16] or semiconductors [17, 18], carbon nanotube functionalization [19], and catalyst pretreatment for carbon nanotube growth [20, 21]. A better insight into the plasma kinetics and plasma surface interactions is essential for many of these applications.

The study of complex plasmas generated from $N_2/CH_4$ gaseous mixtures in capacitively coupled radio-frequency (ccrf) discharges is also a very active field of research. A particular application of this type of discharges is in the simulation of the chemistry of Titan's atmosphere, in laboratory environment. In general, these studies use $N_2/CH_4$ mixtures with low concentrations of methane (< 10%), yet leading to the formation of dust particles analogues to the *tholins*, the solid organic aerosols found in Titan [22]. In Titan's atmosphere, these aerosols are thought to be produced at high altitudes by chemical processes induced by solar radiation and energetic charged particles, both from Saturn's magnetosphere and the solar wind [23]. Since in laboratory plasmas the chemistry is driven by electrons, the characterization of the plasma source and maintenance mechanisms is useful to identify the key processes responsible for particle formation. To this end, a number of different works have been carried out in direct current (dc) discharges [24-28], microwave discharges [29] and their afterglows [30], inductively coupled radio-frequency discharges [31], spark discharges [32], corona discharges [32, 33], dielectric barrier discharges [34] and ccrf discharges [35, 36]. In order to gain insight into the effect of the discharge conditions on the plasma reactivity, but without addressing the additional complexity of dust production, plasmas created from mixtures of $N_2$ with $H_2$, a product of $CH_4$ dissociation, may be used instead.

Low pressure $N_2/H_2$ plasmas have been extensively studied in the last decades, both experimentally [37-49] and theoretically [38, 40, 47-53]. Most of the experimental effort has focused on the synthesis of ammonia [39, 41, 42, 45, 46] and the role of plasma-surface interactions in $NH_3$ production. Different experiments showed that ammonia generation was dependent on the materials of the electrodes and the walls of the plasma reactor, with stainless steel, iron or platinum displaying a catalytic effect larger than other metals or oxides [37, 39, 41, 42, 45, 46, 54, 55]. The prospect of plasma-assisted ammonia synthesis at the industrial scale, with the aim of eventually surpassing the efficiency of the thermochemical Haber-Bosch process, is currently a subject of intensive investigation [54, 56, 57].

On the theoretical side, the first models of $H_2$ and $N_2$ plasmas mainly focused on gas-phase volume reactions and vibrational kinetics [50, 51]. Surface processes were included subsequently by Gordiets *et al.* [52, 53], in order to explain the formation of ammonia, and in the study of the nitriding of iron substrates in this type of discharges [40]. The interaction of the radicals generated in the plasma with the surface of different materials has been analyzed in various works [47, 49, 54, 58], with the aim of determining the relevance of the different surface processes.

The present two-part work studies ccrf discharges, produced in a mixture of $N_2$ with small amounts of $H_2$ (up to 5%), combing both experimental and modelling techniques. This twofold approach is meaningful, considering the complex space-time features of these plasmas, associated with an enhanced volume and surface reactivity. Indeed, by using a model validated against experiment, one can further analyze simulation results to obtain complementary information about the plasma behavior, namely on the rates and paths leading to the production/loss of species. Here, plasmas are produced in a cylindrical parallel-plate reactor surrounded by a grounded metallic cage, corresponding to the PAMPRE experiment [22], with effective coupling powers in the range of 3-13 W and pressures between ~ 0.3 and 1 mbar. In the first part of this study, corresponding to the companion paper [59] hereafter mentioned as I, we presented an experimental characterization of the discharge for the different working conditions, comprising the main electrical parameters, the electron density measured with a resonant cavity technique, and the abundances of stable neutral species and positive ions measured with quadrupole mass spectrometry and infrared (IR) spectroscopy. This paper corresponds to the second part of the study, where simulation results obtained with a hybrid model are presented and compared to the measurements of I. The hybrid model couples a 2D fluid module describing the dynamics of the charged particles with a homogeneous (0D) kinetic module containing a very complete description of the plasma chemistry, both in volume and surface phases. Model validation is based on the measurements of the electrical parameters, the electron density and the fluxes of ions at the wall, as obtained as a function of the effective power coupled to the plasma. Special focus is put on the plasma interaction with the surfaces, highlighting the relevance of secondary electrons in the discharge parameters and the heterogeneous chemistry at the reactor walls, the latter playing a key role in the synthesis of ammonia.

The organization of this paper is the following. Section 2 highlights the most significant model features, to describe both the discharge dynamics and the chemical kinetics. Section 3 presents and discusses the main results obtained, as a function of the effective power coupled to the plasma and at various concentrations of $H_2$. In 3.1 we focus on the electric characterization of the discharge, the plasma electron density and the ionization rate, whereas 3.2 is dedicated to the main results on the neutral and ion species, including ammonia formation. Section 3.3 details the main processes responsible for the kinetics of the discharge. Section 4 concludes.

## 2. Model description

The model employed in this work is based on the ones previously developed by the authors for pure nitrogen [60] and hydrogen [61] ccrf discharges. It is a hybrid code that couples a 2D ($r$,$z$) time-dependent fluid module characterizing the dynamics of charged particles with a very complete 0D kinetic module. The latter includes an extensive set of volume and surface reactions, which are detailed in Table A1 of the appendix.

A detailed description of both the fluid and the kinetic modules can be found in [60]. The fluid module solves the continuity and the momentum transfer equations for electrons, positive ions $N^+$, $N_2^+$, $N_3^+$, $N_4^+$, $H^+$, $H_2^+$, $H_3^+$, $N_2H^+$, $NH^+$, $NH_2^+$, $NH_3^+$ and $NH_4^+$, and negative ions $H^-$ and $NH_2^-$, the electron mean energy transport equations, and Poisson's equation for the rf electric potential. The stationary drift-diffusion approximation is adopted for the particle and energy flux equations, with the introduction of an effective field in the specific case of the ions [59]. The local mean energy approximation is used to obtain the electron transport parameters and the rate coefficients from the space-time electron mean energy profile.

In the fluid module, the electron density $n_e$ and mean energy $\varepsilon$ are obtained from the following continuity equations and boundary conditions:

$$\frac{\partial n_e}{\partial t} = -\frac{1}{r}\frac{\partial\left(r\Gamma_{e_r}\right)}{\partial r} - \frac{\partial \Gamma_{e_z}}{\partial z} + S_e \tag{1}$$

$$\frac{\partial (n_e\varepsilon)}{\partial t} = -\frac{1}{r}\frac{\partial\left(r\Gamma_{\varepsilon_r}\right)}{\partial r} - \frac{\partial \Gamma_{\varepsilon_z}}{\partial z} - \overrightarrow{\Gamma_e}\cdot\vec{E} - S_\varepsilon \tag{2}$$

$$\left.\frac{\partial n_e}{\partial r}\right|_{r=0} = \left.\frac{\partial n_e}{\partial z}\right|_{r=0} = 0 \tag{3}$$

$$\left.\frac{\partial (n_e\varepsilon)}{\partial r}\right|_{r=0} = \left.\frac{\partial (n_e\varepsilon)}{\partial z}\right|_{r=0} = 0 \tag{4}$$

where $\boldsymbol{E}$ is the electric field, $\boldsymbol{\Gamma}_{e/\varepsilon}$ is the electron particle/energy flux, $S_e$ is the net production rate of electrons, and $S_\varepsilon$ is net power density lost by the electrons due to elastic and inelastic collisions.

This set of equations for the plasma (slow) electrons is complemented by a single-beam model for fast electrons generated by secondary emission at the walls, following the one proposed for dc discharges in [62], where a detailed formulation can be found. The model considers a beam of monoenergetic fast electrons travelling in a straight line perpendicular to the emitting surface, and solves the evolution of the beam flux and energy along its path. The fast electrons are assumed to be created at the surfaces in contact with the plasma (both electrodes and the surrounding cage) by impingent positive ions, and their initial energy is set to 1 eV (numerical tests showed that simulation results are not sensitive to variations of this value within 50%). At the surface, the fluxes $j$ of fast electrons (signaled with subscript $f$) and positive ions (+) are related by:

$$j_f = -\gamma\, j_+ \tag{5}$$

where $\gamma$ is the secondary electron yield. This parameter is adjusted to fit the experimental trends of the electron density as a function of the effective power coupled. The coupling between fast and slow electrons is as follows. As fast electrons traverse the sheath, they are accelerated by the electric field. When their energy is larger than the ionization threshold, they can be responsible for the ionization of neutrals and the production of new beam electrons, thus increasing its flux and reducing its energy. The beam is fully diluted into the plasma when the fast electrons reach the negative glow. There, ionizations produce slow electrons instead, and the beam energy is progressively lost to inelastic collisions (both ionizations and excitations) until it lies below the lowest excitation threshold, in which case the beam electrons become part of the (slow) plasma electrons. The particle and energy balance equations for the beam (fast) electrons are written as in [63]:

$$\frac{\partial n_f}{\partial t} = -\nabla_\perp \left(n_f u_{\perp f}\right) + S_f \tag{6}$$

$$\frac{\partial \left(n_f \varepsilon_f\right)}{\partial t} = -\nabla_\perp \left(n_f u_{\perp f} \varepsilon_f\right) + n_f u_{\perp f} E_\perp - S_f^\varepsilon \tag{7}$$

where $n_f$ is the density of fast electrons, $u_f$ is the drift velocity, $S_f$ and $S_f^\varepsilon$ are the source and loss terms for the particle and energy balance equations, respectively, and $\perp$ represents the vector-components perpendicular to the walls. Equations (6)-(7) are solved by numerical integration over a discrete grid that enhances the spatial description of the space-charge sheath (typical cell-sizes of 0.5 mm). Its solution coupled to the plasma equations (1)-(2) is particularly relevant to self-consistently account for changes in the electric field and the plasma source terms, leading to calculated values of the electron density much closer to the measurements at high powers.

The 0D kinetic module solves the two-term homogeneous and stationary electron Boltzmann equation for the main plasma species $N_2$, $H_2$, $NH_3$, N, H (accounting for inelastic collisions from ground-state molecules and atoms, and inelastic and superelastic collisions involving vibrationally excited states), and the set of (average) rate balance equations for 78 additional neutral species. These include the 15 vibrational states of $H_2$(X) and the 46 of $N_2$(X), 8 electronic excited states (7 for $N_2$ and 2 for N), 5 adsorbed species (N(s), H(s), NH(s), $NH_2$(s) and the free surface sites F), and other molecules and radicals. The steady-state solution of the 0D rate balance equations is computed after every few rf cycles (typically 200) solved by the fluid module, but contrary to [60], this is done by means of a stiff solver (LSODE [64]) that calculates the time evolution of the system to ensure the stability of the solution, instead of directly solving the stationary form of the species balance equations.

For the kinetic scheme, we have combined the reactions from our previous models for $N_2$ [60] and $H_2$ [61], complemented by additional processes involving both species, taken mostly from [52] and [65]. The rates for electron-impact processes are calculated using up-to-date cross section sets from the IST-Lisbon [66, 67] ($N_2$ [68-70], N [71, 72], $H_2$ and H [61, 73-82]) and Hayashi ($NH_3$) [83, 84] databases, available at LXCat. The vibrational processes involving both $N_2$ and $H_2$ are taken from [52]. In the case of the e-V processes in $N_2$, the IST-Lisbon database contains only the cross sections for the excitation from the ground-state to the first 10 vibrational levels. The rate coefficients for $v \geq 1$ are calculated from the former using the same scaling law as in [71]:

$$k_{v,v+n}(T_e) = \frac{k_{0,n}(T_e)}{1 + av}, \quad 1 \leq n \leq 10 \;\; ; \quad 0 \leq v \leq 45 \tag{8}$$

and taking $a = 0.15$.

The surface chemistry model, describing the multi-step formation of ammonia at the reactor walls, is mainly taken from [47]. The reaction set has been extended to include the dissociative adsorption of molecules following [58]. The coefficients given in [47] correspond to a stainless steel surface, while the PAMPRE experiment features a mixing of stainless steel (electrodes) and aluminium alloy (cage) walls. As noted in [58, 85], the reaction coefficients for hydrogen and nitrogen can vary significantly depending on the surface material, and in particular, a reduction of an order of magnitude is expected for $Al_2O_3$, which could be present in our experiment. Thus, the set of reaction coefficients has been tweaked towards lower reactivity to account for this and better fit the experimental $NH_3$ abundances.

The rates due to the flow of species in and out of the chamber are calculated similarly to [86]. The rate at which precursors enter the chamber is given by:

$$R_{in} = 4.49 \times 10^{17} Q_{in}/V \text{ cm}^{-3} \text{ s}^{-1} \quad (9)$$

where $Q_{in}$ is the inlet flow in sccm and $V$ is the reactor volume in $cm^3$. To balance the possible changes in the number of particles in the reactor due to the plasma kinetics, the rate at which a species X leaves the reactor is calculated as:

$$R_{X,out} = R_{in} \frac{n_X}{n_{T,0}} \quad (10)$$

where $n_X$ is the density of species X and $n_{T,0}$ is the total particle density in the chamber before discharge ignition. This ensures the conservation of pressure throughout the simulation.

As mentioned, the solution of the model is obtained by alternate iterations of the fluid and the kinetic modules. The solution to the fluid module gives the effective power coupled to the plasma ($W_{eff}$) and the self-bias voltage ($V_{dc}$), along with the space-time evolution of the charged-particle densities, fluxes and electron-impact reaction rates; the electron mean-energy and energy-flux; and the space-charge electric field. From these results one can calculate the space-time averaged reaction rates involving charged species, which are used in the kinetic equations of the 0D module to obtain the volume and time-averaged densities and reaction rates for the neutral species. These newly calculated densities are used as input to solve the electron Boltzmann equation, and the resulting updated electron transport parameters and rate coefficients are then employed in the following iteration of the fluid module.

To reduce the convergence time of the code, a preliminary solution is quickly obtained by using a large timestep. This leads to a somewhat inaccurate treatment of the transport, but nevertheless constitutes a very good starting point for the final solution, which is then reached by reducing the timestep to an adequate value. Overall, runtimes can be reduced by up to 70% by using this method (for convergence criteria imposing relative errors for the calculated quantities below $10^{-3}$, typical runtimes of 24–100 hrs are obtained in an Intel-Xeon E2697 V2 (2.7 GHz) CPU).

# 3. Results and discussion

This section presents and discusses the main results obtained in the simulations, comparing them with the measurements available (see I). Similarly to the experiments, the working conditions are for pressures in the range of 0.36-0.92 mbar (gas flows between 20–55 sccm) and coupled powers in the range of 1-15 W, considering $N_2$-$H_2$ mixtures with $H_2$ fractions up to 0.05.

Note that the model uses as input parameter the rf-voltage $V_{rf}$ applied to the driven electrode, yielding $W_{eff}$ as self-consistent result [59,60]. However, we have preferred to use the latter in the comparisons with the experiment, as in [59], considering that $V_{rf}$ was not measured at the electrode [59] and that $W_{eff}$ provides direct information about the power available for plasma excitation.

## *3.1 Electrical parameters and electron density*

The results for the electrical parameters of the discharge are shown in Figure 1. Here, calculated and measured values of the electron density and the self-bias voltage are displayed as a function of the coupled power, for a mixture with 0.05 $H_2$ fraction at two different pressures (0.50 and 0.92 mbar).

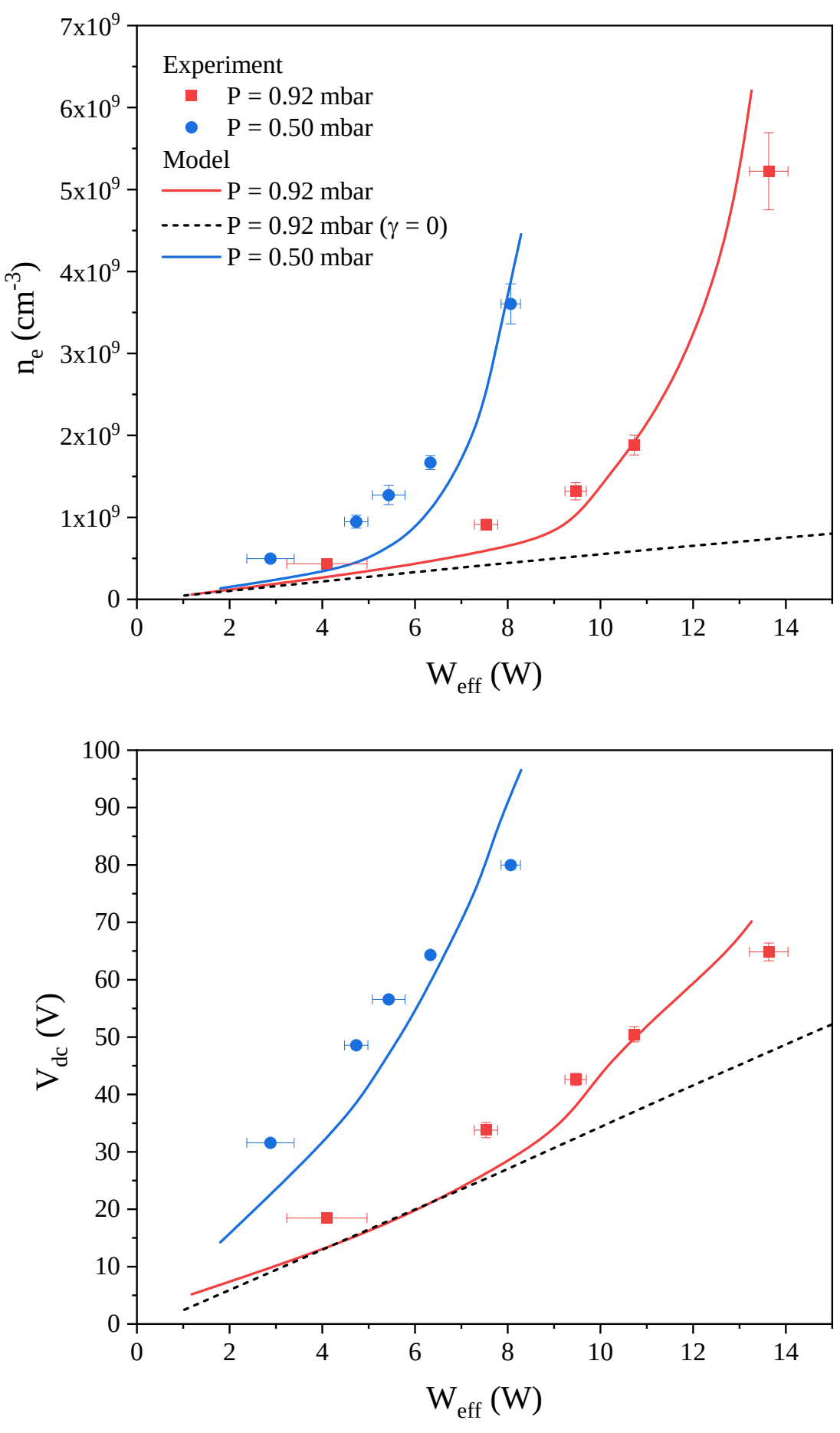

**Figure 1.** Simulations (curves) and measurements (points) for the electron density (upper panel) and the self-bias voltage (lower panel), as a function of the power coupled to the plasma, for a mixture with 5% $H_2$ at the following pressures (in mbar): 0.5 (blue circles and lines); 0.92 (red squares and lines). The dotted curve was obtained at 0.92 mbar, neglecting the electron production by secondary emission.

Both the experiments and the simulations show that the electron density increases linearly at low power values, exhibiting an exponential growth at higher powers. An increase in the pressure (hence in the gas density) leads to a displacement of the curve $n_e(W_{eff})$ towards higher effective powers as expected, without changing its global behavior. In general, model predictions underestimate the electron density, and this is more evident at the lower pressure where a factor of two difference can be found with respect to the experimental values. At high power, the exponential growth is entirely due to the secondary electron emission. If the contribution from the secondary electrons is not included, the electron density exhibits a purely linear growth with the power, as shown in fig.1. As mentioned, the secondary electron yield $\gamma$ is adjusted to fit the highest $n_e$ value for each pressure, resulting in $\gamma$ = 0.09 at 0.50 mbar and $\gamma$ = 0.06 at 0.92 mbar. These values are in the typical range for this type of discharges [87, 88], and the decrease in the yield with increasing pressure can be justified by the lower energy of the heavy particles reaching the walls [88].

As expected, the self-bias voltage increases with the coupled power and decreases with the pressure, in direct relation with the intensity of the ion current density at the reactor walls. Model predictions for $V_{dc}$ are in good qualitative agreement with the measurements, reproducing the behavior of this parameter with changes in both $W_{eff}$ and $p$. The calculated values underestimate the measurements at low power, but this difference is reduced as the coupled power increases. The model behavior is mostly linear, somewhat deviating from that trend when the secondary electron emission becomes significant.

In the experimental part (see I), a weak upwards trend was observed for the electron density when increasing the hydrogen content in the precursor mixture at constant power. The corresponding model results are shown in Figure 2.

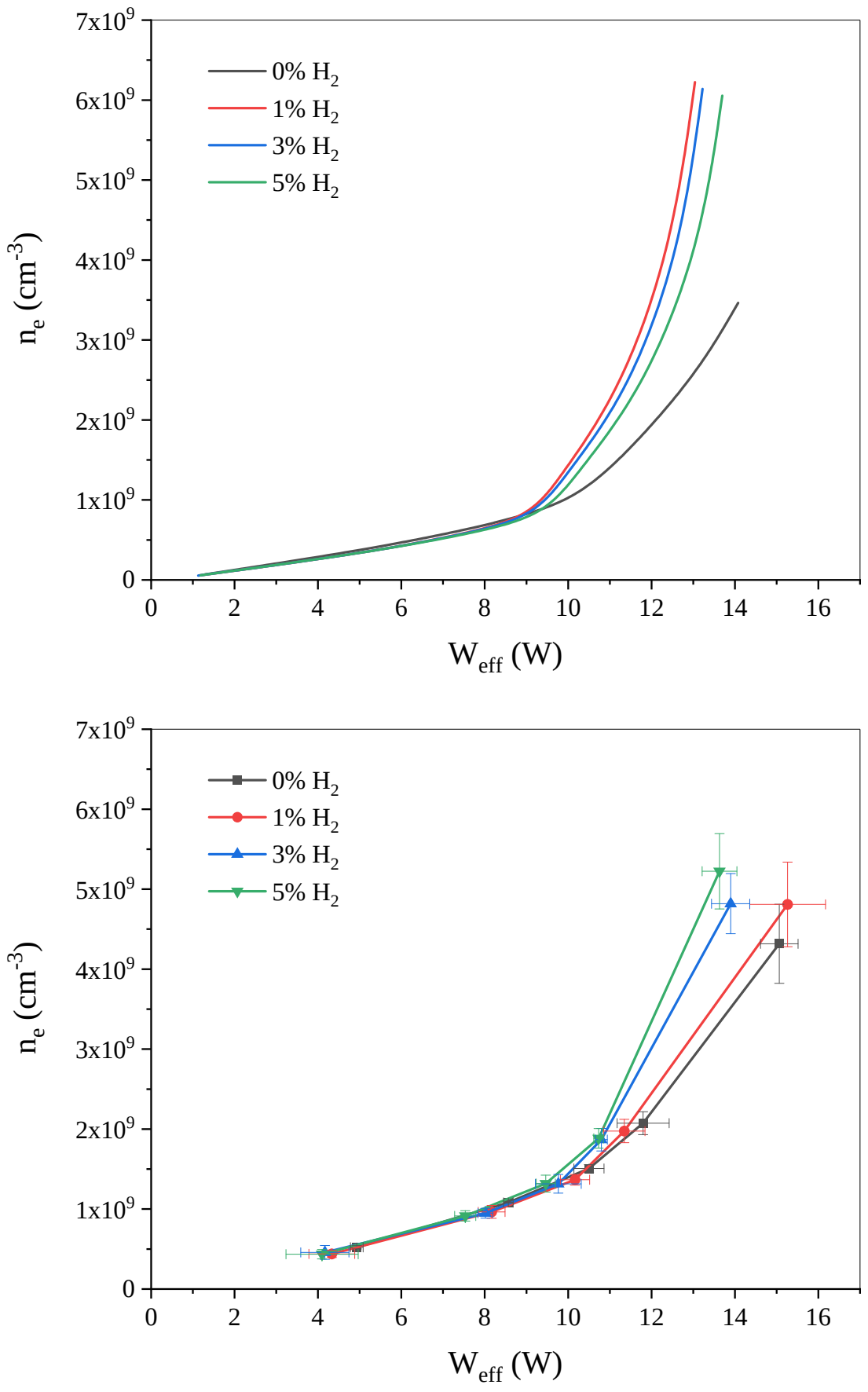

**Figure 2.** Electron density as a function of the power coupled to the plasma, at a pressure of 0.92 mbar and for different $H_2$ fractions in the gas mixture (in percentage): 0 (black), 1 (red), 3 (blue), 5 (green). Upper panel: simulations. Lower panel: experiments.

Here we observe a behavior different from the experiments. The lowest electron density is also obtained for the pure $N_2$ discharge, however, instead of a smooth increase with the $H_2$ fraction, a sharp jump in $n_e$ is found for the 1% $H_2$ mixture. Moreover, further addition of hydrogen results in a steady decrease of the electron density, although the value at 5% is still above the pure $N_2$ result. This discrepancy might be due to several factors. First, because of the exponential growth behavior of $n_e$ at high powers, a small error in the determination of $W_{eff}$, either in the experiments or in the model, could be enough to change the observed behavior. Second, this trend is specifically observed at higher powers, where ionization is dominated by the secondary electron production (see the discussion below). Since the model for fast electrons is rather simplified, fine variations such as this one should be taken with caution.

Typical electron energy distribution functions (EEDF, $f(u)\sqrt{u}$, normalized so that $\int_0^\infty f(u)\sqrt{u}du = 1$) obtained for the slow electrons are shown in Figure 3, for pure $N_2$ and 5% $H_2$ mixtures, at $p$ = 0.92 mbar, $W_{eff}$ = 5 W and 11.5 W, and for two different positions along the discharge axis, corresponding to the plasma bulk (half-way between the electrodes) and within the space-charge sheath. Under low power conditions, slow electrons are responsible for most of the ionization processes and fast electrons are less relevant. In this case, an observable difference is found between the distributions in the bulk and in the sheath, the latter showing a significantly populated tail up to 12 eV, while the former drop more abruptly. Further examination of the results reveals that, starting at energies > 3 eV (beyond the peak of the distribution), the EEDFs in the sheath are ~10–100 times larger than in the bulk, yielding significant differences in the efficiency of electron-impact excitations and ionization in both regions.

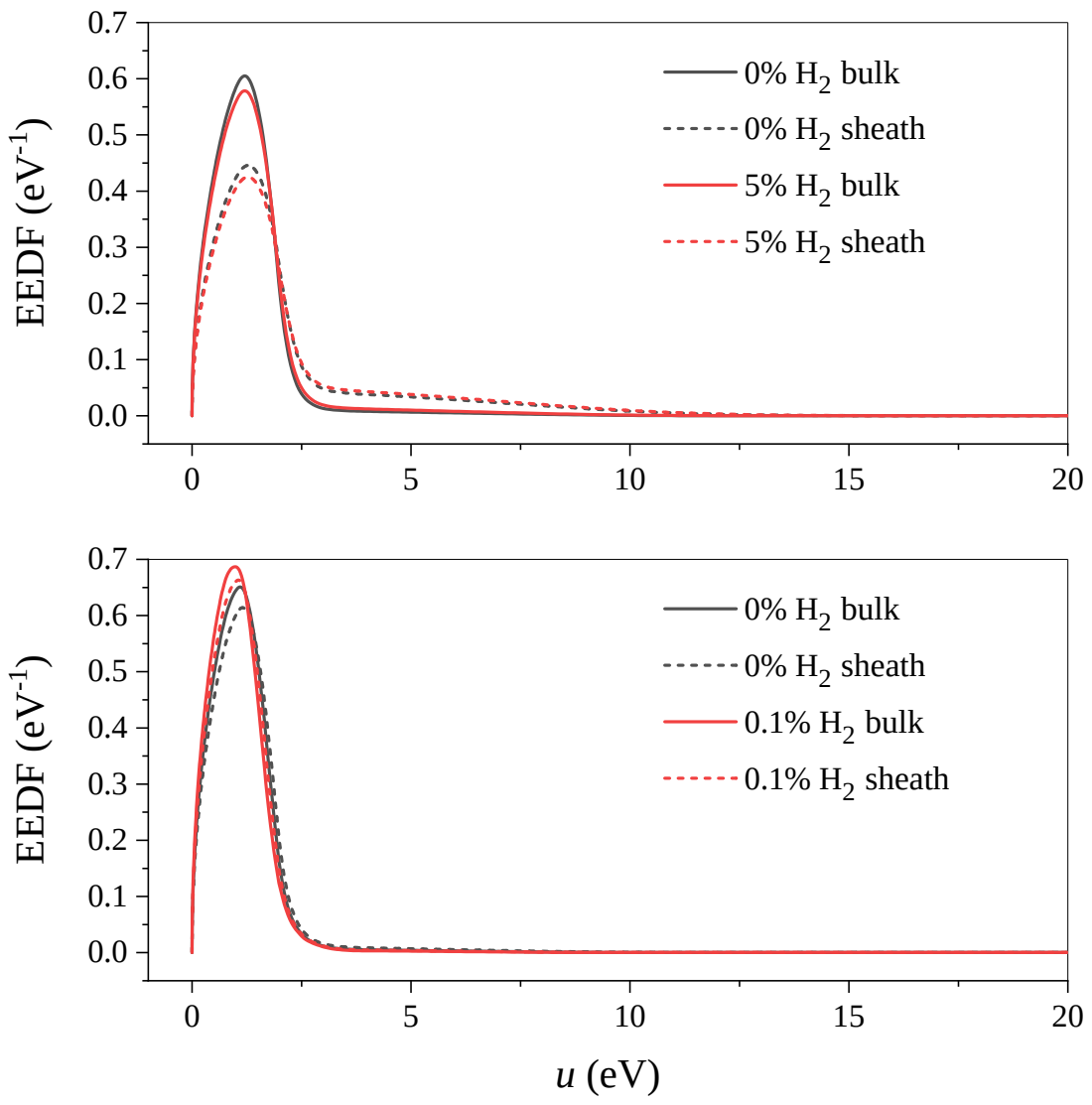

**Figure 3.** Calculated EEDFs of the plasma (slow) electrons, at 0.92 mbar and for two different positions along the discharge axis, corresponding to the plasma bulk (solid lines) and within the space-charge sheath (dashed). The upper panel is for a low-power of 5 W, in pure $N_2$ (black curves) and 5% $H_2$ mixtures (red); the lower panel is for a high-power of 11.5 W, in pure $N_2$ (black curves) and 0.1% $H_2$ mixtures (red).

Much smaller differences are found when comparing the EEDFs for different gas compositions, at low power. In this case, and for both bulk and sheath regions, the addition of $H_2$ to the mixture results in only a slight increase of the EEDF tail, with the corresponding decrease of its body , thus leading to minor changes in the electron-impact excitation and ionization rates (in agreement with the results in Figure 2). Under high power conditions, the EEDFs for slow electrons are very similar in different regions and at different gas compositions. A small increase in the body of the distributions is observed in the bulk region or for 0.1% $H_2$ mixtures, translating into a ~10% decrease in the ionization rate, contrary to what one could expect from Figure 2. Indeed, the results therein revealed a non-monotonic variation of the electron density with the percentage of $H_2$ in the mixture, probably associated with the production of fast secondary electrons, which would control electron-impact ionization phenomena.

The influence of the fast electrons in the plasma can be visualized by comparing the time-averaged densities, mean energies and ionization rates for both slow/fast electrons. The results are shown in the contour plots of Figures 4 and 5, obtained for a pressure of 0.92 mbar, a high-power of 11.5 W, and $H_2$ percentages of 5% (Fig. 4) and 0% / 0.1% (Fig. 5). The figures for fast electrons display the added contributions of three beams, originated at the powered electrode, the grounded electrodes and the surrounding cage.

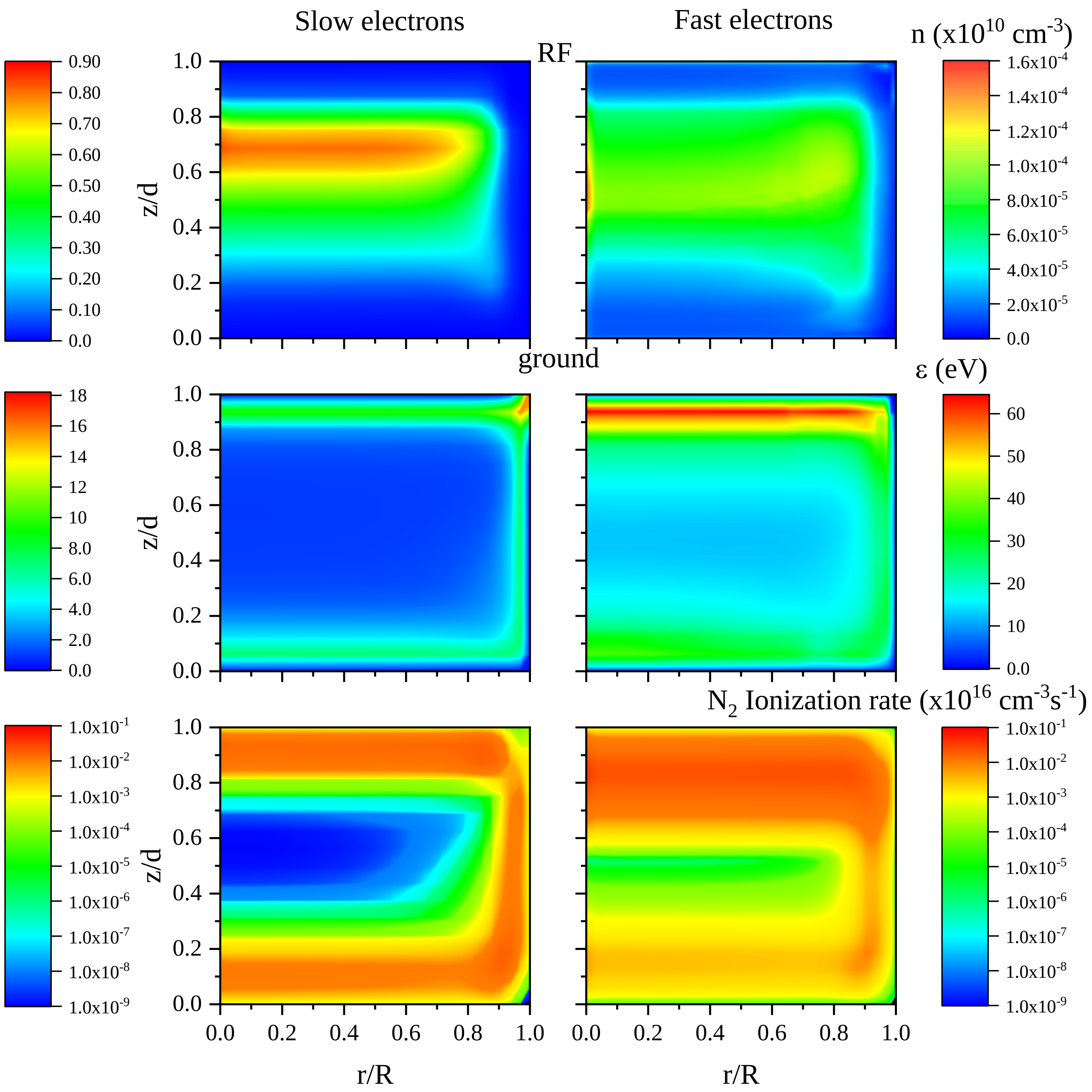

**Figure 4.** Calculated contour plots of time-averaged densities, mean energies and $N_2$ ionization rate contributions for slow (left) and fast (right) electrons, for a 5% $H_2$ mixture at 0.92 mbar and 11.5 W. The spatial scales are normalized to the discharge radius *R* and the interelectrode distance *d*. Note the different colour scales in the slow and fast electron plots. The labels "RF" and "ground" indicate the position of the powered ($z/d$ = 1) and grounded ($z/d$ = 0) electrodes, respectively.

The density and energy panels in Figure 4 highlight the differences between both groups of electrons. The results for the density show that both, fast and slow electrons, are present in the discharge bulk, with slow electrons exhibiting a maximum density closer to the powered electrode ($z/d$ = 1) and fast electrons distributed rather homogeneously at density values ~4 orders of magnitude lower. Indeed, for most of the working conditions considered, the fraction of fast electrons is ~0.01–0.02 % of the total electron density. The results for the energy of fast/slow electrons are essentially the opposite of what is observed for their densities. While slow electrons typically reach mean energies of only ~10 eV due to collisional heating, the fast electron beam can be accelerated up to ~60–80 eV by the electric field, within the sheath close to the powered electrode, and easily sustain 20–40 eV elsewhere, well above the ionization threshold.

The effect of both densities and energies is compounded in the ionization rate. For the comparison, we have chosen to show in the lower panels of Figure 4 the contribution of both groups of electrons to the $N_2$ ionization rate. As expected, the main ionization region is found in the sheaths, for both groups of electrons and with similar intensities, but a difference of a few orders of magnitude is encountered in the discharge bulk. While for the slow electrons the ionization in the center of the plasma is negligible, for the fast electrons it becomes a relevant source of charged particles. The result is a significant increase in the volume-averaged electron density, as shown previously in Figures 1-2, and a substantial change in the plasma ion chemistry.

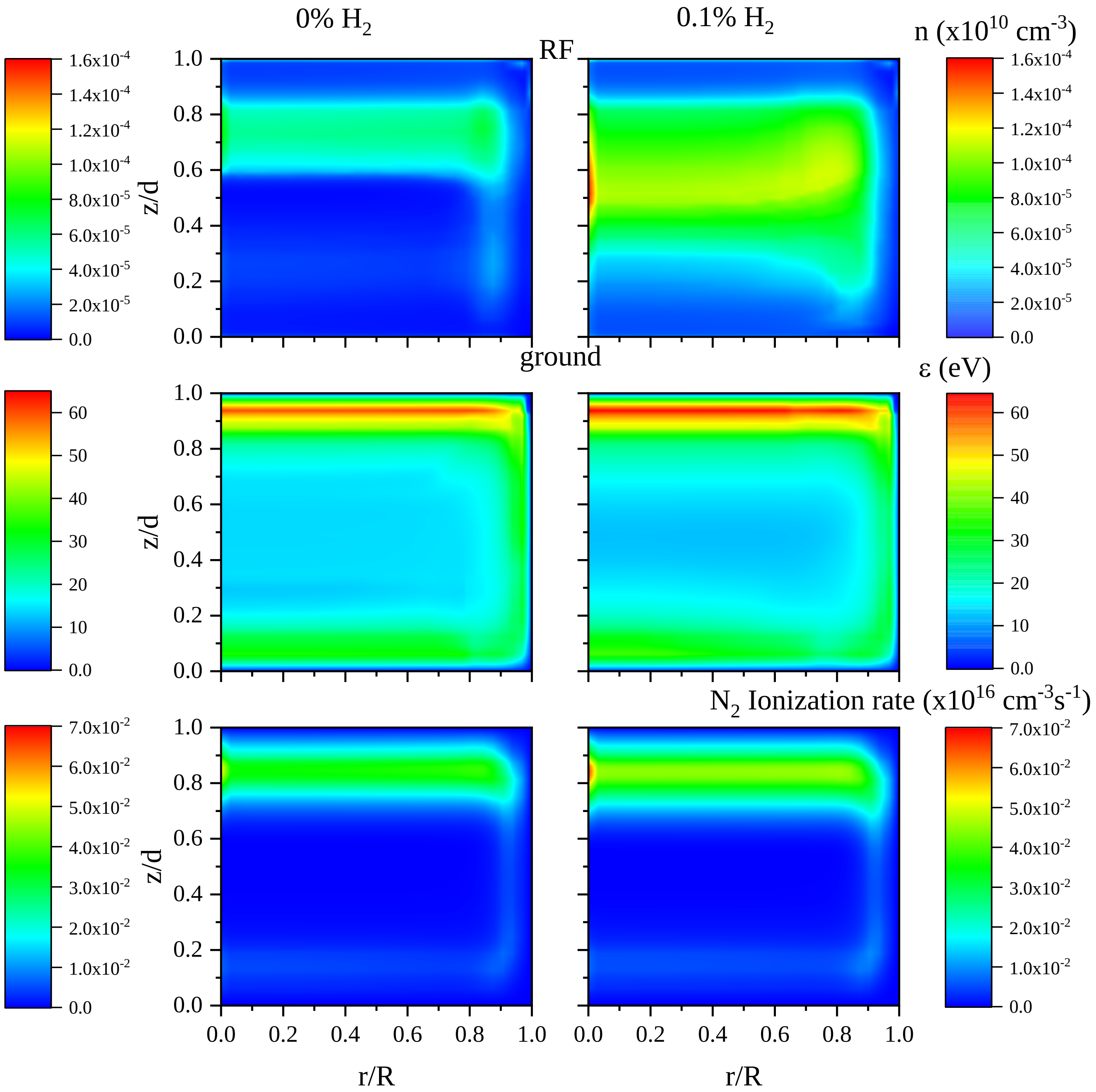

**Figure 5.** As in Figure 4, but only for fast electrons in mixtures with 0% $H_2$ (left) and 0.1% $H_2$ (right).

The variations in the contribution of fast electrons to ionization, at high-power and various mixture compositions, can be analysed in Figure 5. Here, we observe that the slight introduction of $H_2$ yields a mild increase in the mean energy and a large increase in the electron density and ionization rate, the latter becoming ~20% higher, in agreement with the electron density jump observed in Figure 2 when the mixture composition goes from 0% to 0.1% $H_2$. The major cause for this effect is related with the reduction in the ionization threshold for hydrogenated gas mixtures, which is then set at 10.2 eV (corresponding to $NH_3$) instead of the 15.4 eV (for $N_2$) required to ionize pure nitrogen discharges. This ionization uprise is probably overestimated by the model adopted for the fast electrons beam, where the ionization threshold controls the beam-growth in the sheath and its decline in the bulk. The effect is naturally mitigated with the continuous increase in the $H_2$ content (due to the enhanced contribution for the ionization of both $N_2$ and $H_2$, with similar thresholds), as it is also observed in Figure 2.

### *3.2 Ion and neutral densities and fluxes*

The behavior of charged particles can also be analyzed through the ion fluxes or density profiles. Figure 6 compares the major ion fluxes perpendicular to the walls, measured with a quadrupole mass spectrometer (QMS) (see paper I) and predicted by the model within a rf cycle (average value) along the *z* axis at $r = R$. The results reveal that the most abundant ion is $N_2H^+$, whereas the fluxes for $N_2^+$ and $NH_4^+$ are found to be significantly lower. Simulations show a similar behavior for all positions along the *z* axis, but the values of the fluxes are closer to the measured ones in the sheath region where the electron mean energy and the ionization rate are higher (note that the QMS measurements are performed approximately halfway between electrodes). This result, together with the lower electron densities predicted at low powers, confirms that the model predictions underestimated the ionization rate in the bulk.

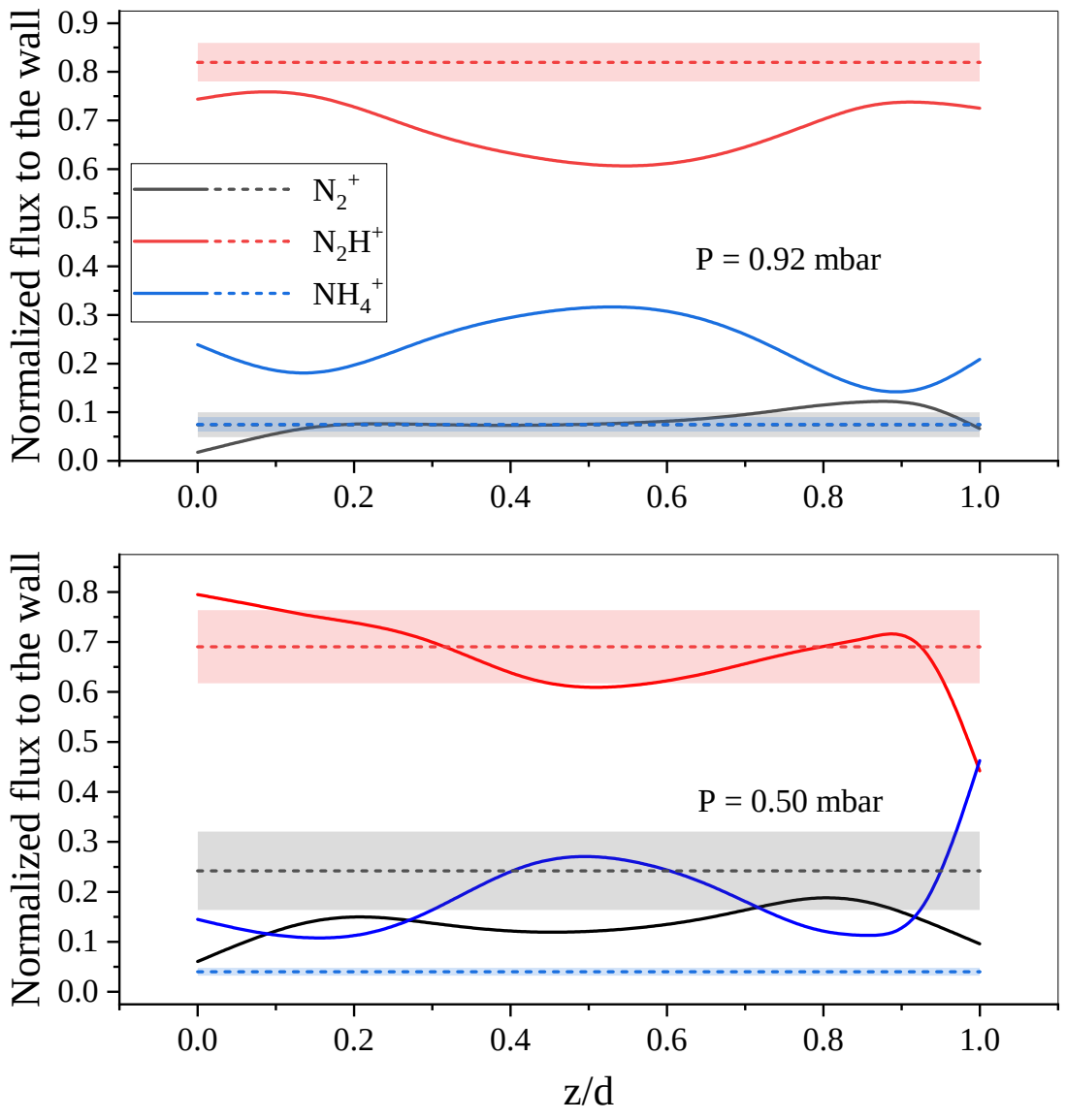

**Figure 6.** Normalized ion fluxes to the wall at $r=R$, as a function of the normalized distance to the grounded electrode, for $N_2^+$ (black lines), $N_2H^+$ (red) and $NH_4^+$ (blue), for a 5% $H_2$ mixture at two different pressures: 0.92 mbar, 10 W (upper panel) and 0.5 mbar, 6 W (lower panel) . The solid lines are model simulations and the dashed lines and shades represent the value measured with the QMS, at approximately $z=d/2$, and its error.

Figure 7 shows the measured and simulated ion fluxes at the wall as a function of the $H_2$ fraction in the mixture. Overall, the behavior of the ions is well reproduced, specifically the $NH_4^+$ and $N_2H^+$ growth and $N_2^+$ decrease as hydrogen is injected. The normalized fluxes of the four major ions in the discharge ($N_2H^+$, $NH_4^+$, $N_2^+$, and $NH_3^+$) are very well reproduced, however, certain other somewhat minor species show significant quantitative discrepancies. This is the case for $NH^+$, $NH_2^+$ and $N_4^+$. This discrepancy holds for the rest of the comparisons shown below. Due to this, we have decided to only show the results for the major ions henceforth. A small difference is also observed in the $NH_3^+$ trend in what is otherwise a good quantitative agreement between model and experiment. The model overestimation of this ion could possibly be related to the overestimation of the $NH_3$ abundance for mixtures with < 5% $H_2$, as will be shown below (see Figure 10). The experimental detection of the pure hydrogenic ions $H_2^+$ and $H_3^+$ requires a different set of QMS acquisition parameters and cannot be directly related to the rest of the ions (see paper I). They are nevertheless included in the figure to show the trend against the $H_2$ fraction, but the experimental values displayed there do not represent their normalized ion flux and therefore should not be compared with the model results.

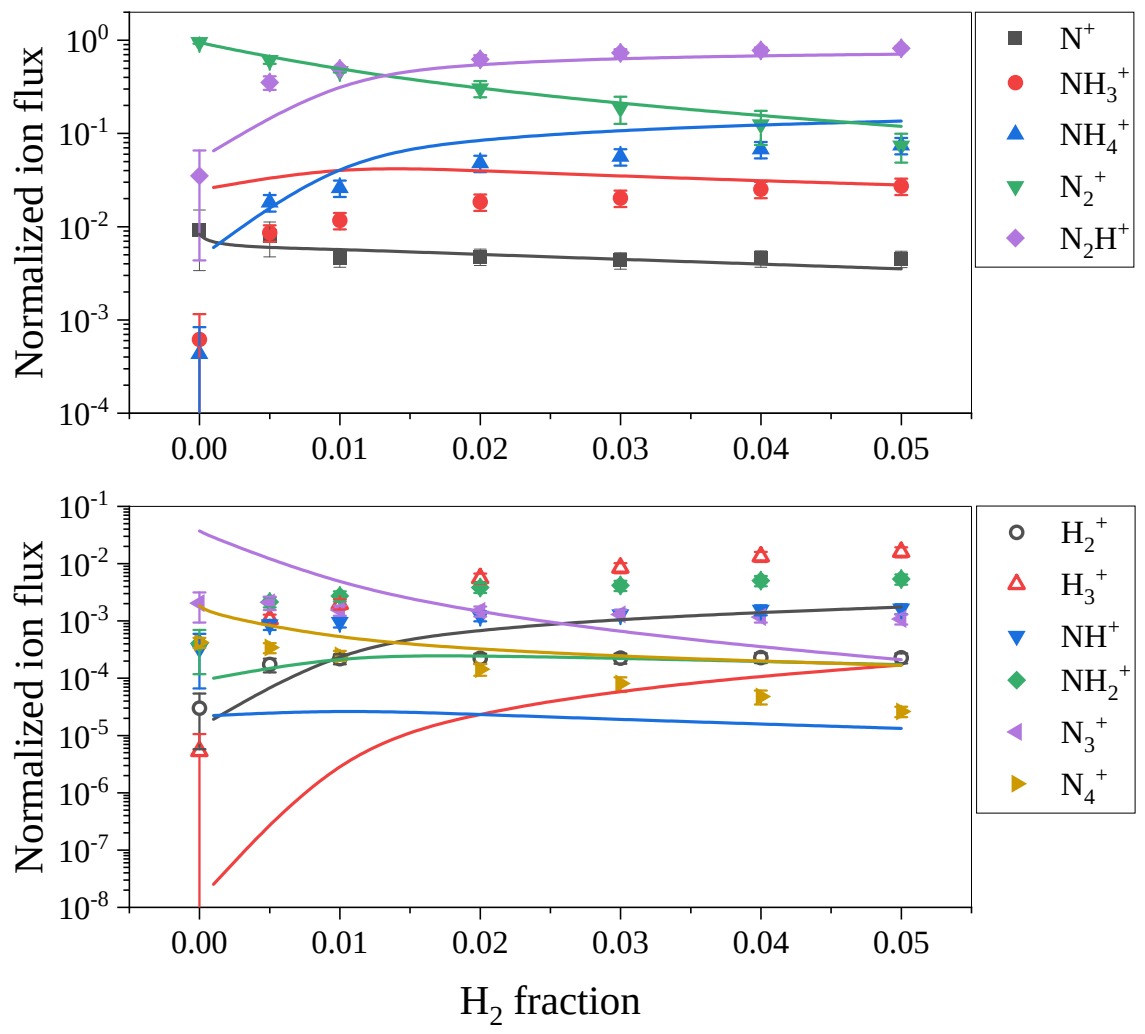

**Figure 7.** Calculated normalized ion fluxes to the wall as a function of the $H_2$ content in the mixture for a discharge at 0.92 mbar and 10 W (lines), and comparison with the experimental measurements (dots). Upper panel: major ions. Lower panel: minor ions. Open symbols for $H_2^+$ and $H_3^+$ correspond to uncalibrated measurements and do not represent the ion flux (see text).

The evolution of the major ion fluxes with the pressure is depicted in the upper panel of Figure 8. Smaller changes to the distribution are observed in this case, both in the model and in the experiments. However, the trends predicted by the model are smoother than the ones observed experimentally, which is evident in the case of $N_2^+$ and $NH_4^+$. Nevertheless, the global trends are adequately reproduced and the evolution of the rest of the major ions is satisfactorily captured.

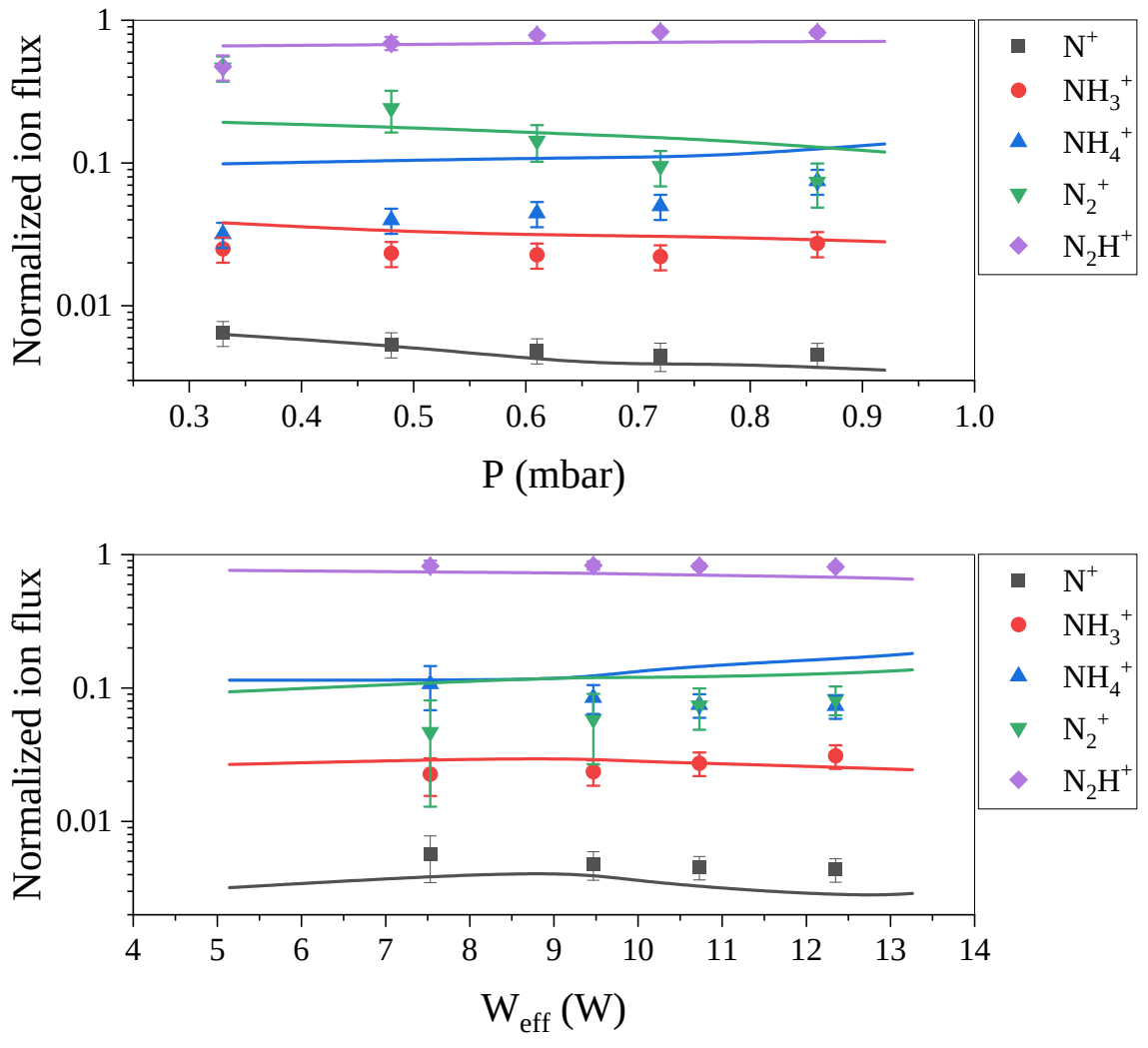

**Figure 8.** Model (lines) and experimental (dots) normalized ion fluxes as a function of the pressure (upper panel) and the power coupled to the discharge (lower panel). Discharge conditions are fixed at 5% $H_2$, and 10 W and 0.92 mbar respectively. Only major ions are shown for brevity.

Lastly, the evolution of the major ions with the coupled power is shown in the lower panel of Figure 8. In this case, the comparison between model and experiment is worse than in the previous ones. The model and the experiment show small changes in the distribution, but the trends for these are opposite in most of the cases. The only ions for which the model shows a similar trend to the experiments are $N_2^+$ and $N_2H^+$, and only in the case of $N_2H^+$ do we also have a quantitative agreement on the flux. This may be due to the heavy growth in the total ion density associated with an increase in power, which could make the comparison of normalized ion fluxes a bit less accurate than in previous cases. On a related note, higher power values also involve a heavy influence of the secondary electrons in the results, which, as mentioned before, is somewhat simple and perhaps unable to correctly account for the ion chemistry in the discharge.

The flux of ions to the wall does not give a complete idea of the plasma ion composition. Figure 9 (upper panel) presents a comparison between the ion fluxes and the volume-averaged ion densities predicted by the model in the same conditions as the upper panel of Figure 6. The results in this figure show that while the largest ion flux corresponds to the $N_2H^+$ ion, the volume-averaged densities are dominated by $NH_4^+$, followed by $N_2H^+$, with $N_2^+$ appearing in concentrations up to an order of magnitude lower than the former. Negative ions $H^-$ and $NH_2^-$ , which are trapped in the discharge by the plasma potential and thus have no flux to the walls, are shown to also be present in significant amounts (~1% of the total charge) in the plasma bulk.

The cause of the differences between the volume-averaged densities and the fluxes to the wall becomes clear after analyzing the axial distribution of the three major ions, depicted in the lower panel of Figure 9. $N_2^+$ is found to be present in significant amounts only in the plasma sheaths, while its concentration drops in the bulk. This behavior is mirrored by $N_2H^+$, albeit with higher concentrations at all positions and, particularly, close to the electrodes. Finally, $NH_4^+$ shows the opposite trend, being mostly absent in the plasma sheaths and quickly growing in the bulk, becoming the dominant ion in this region. Note that the difference in the results for the ion average-densities and wall-fluxes justifies a careful comparison between model predictions and mass-spectrometry measurements, since the latter provide direct information only on the *local* wall-fluxes (see I).

This differences in the ion composition close to the reactor walls and in the discharge bulk can be justified by recalling that the space-charge sheaths are the regions with the higher ionization rates, and especially the one near the rf-driven electrode (see lower panel in Figure 4), due to the strong variation of the electric field in these regions and the presence of energetic secondary-electrons coming from the surfaces. In particular, ionization of $N_2$ triggers a chain of ion-molecule reactions leading to the formation of $NH_4^+$ (see section 3.3):

$e + N_2 \rightarrow N_2^+ + 2e$ (e28)

$N_2^+ + H_2 \rightarrow N_2H^+ + H$ (im6)

$N_2H^+ + NH_3 \rightarrow NH_4^+ + N_2$ (im24) ,

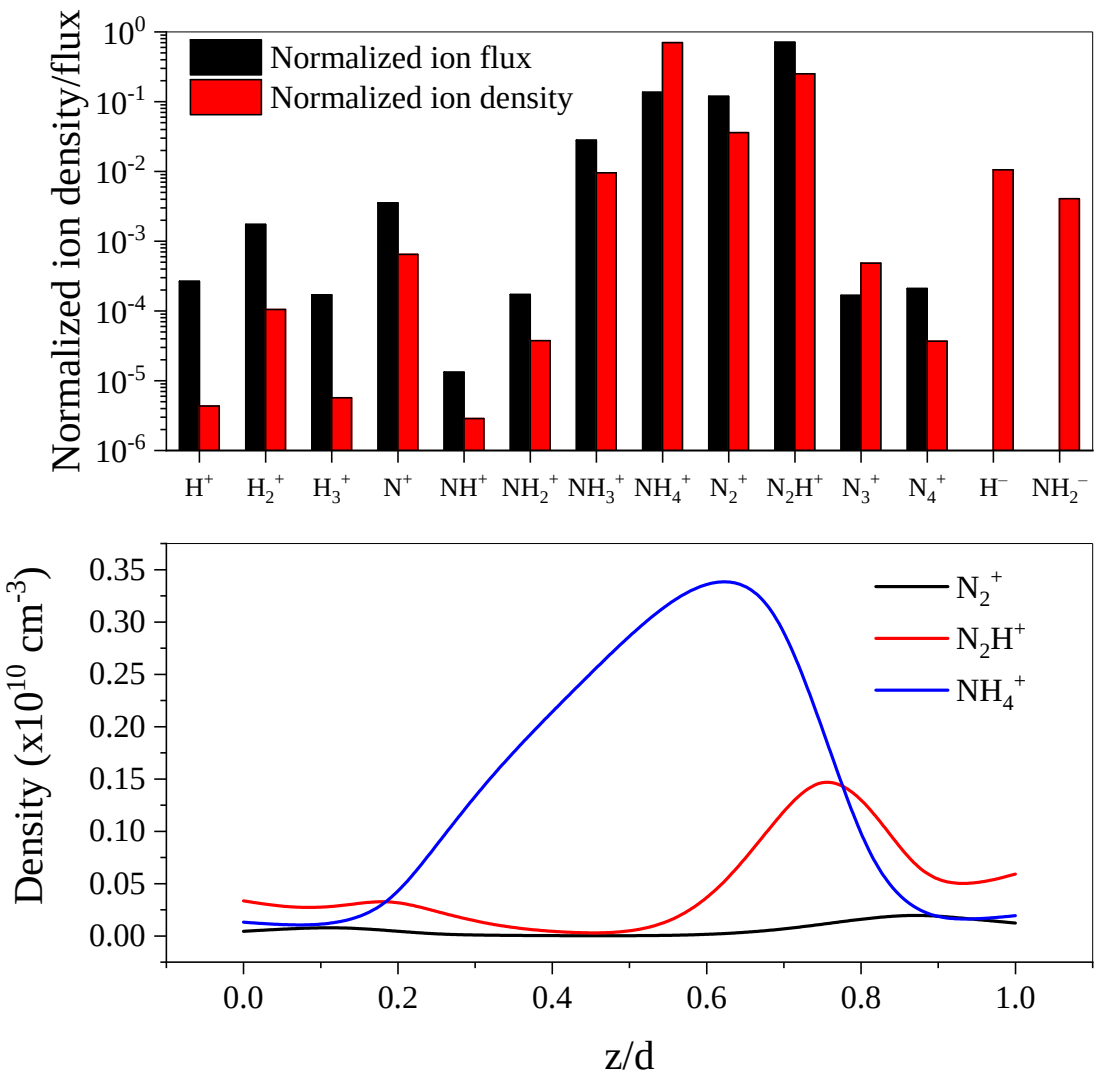

**Figure 9.** Calculated ion densities and fluxes for the same conditions as the upper panel of Figure 6. Upper panel: normalized ion fluxes at the wall (black bars) and volume-averaged normalized ion densities (red bars). Lower panel: absolute densities for the three most abundant ionic species, at the axis of the discharge ($r$ = 0) as a function of the distance to the grounded electrode (same color codes as in Figure 6).

and because both ion-molecule reactions are very efficient, the relative abundance of these three ions is controlled by the efficiency of the ionization reaction (e28). In high ionization regions, the equilibrium is displaced towards the formation of $N_2^+$, resulting in $N_2H^+$ being the major ion reaching the surface, while in the center of the discharge, where little ionization takes place, most of these ions are converted into $NH_4^+$, if $NH_3$ is present in sufficient amounts.

The abundance of ammonia plays a key role in the ion plasma chemistry, and is primarily dependent on the balance between electron impact dissociation of the precursors and the subsequent surface recombination. A study of the influence of the discharge working conditions on the formation of $NH_3$ is shown in Figure 10. The abundance of ammonia is found to increase with the fraction of $H_2$, as expected, and with increasing power, yet decreasing when the pressure increases. These trends are qualitatively reproduced by the model, but simulations and measurements disagree in the details of the variation of $NH_3$ concentration with pressure and power. This is especially evident at low pressure, where model predictions underestimate the $NH_3$ concentration by a factor of ~2.

The dependence of the ammonia concentration with pressure is most intriguing, since the model predicts an increase in the absolute density of $NH_3$ with pressure, whereas the experiments show a decrease in the absolute density for pressures higher than 0.66 mbar (see I). This hints at the presence of a mechanism not included in the model, perhaps related to the vibrational kinetics of $N_2$, such as vibrationally enhanced dissociative adsorption [89]. Simulation results predict a weak variation of the (volume-averaged) vibrational distribution function (VDF) of $N_2$ with the pressure, but the scenario could change in a spatially-resolved chemistry model coupled to the fast electrons, with enhanced electron-impact excitation and dissociation close to the walls. The effective surface reactivity, which is kept constant in the model, could also change with pressure thus affecting the concentration of $NH_3$.

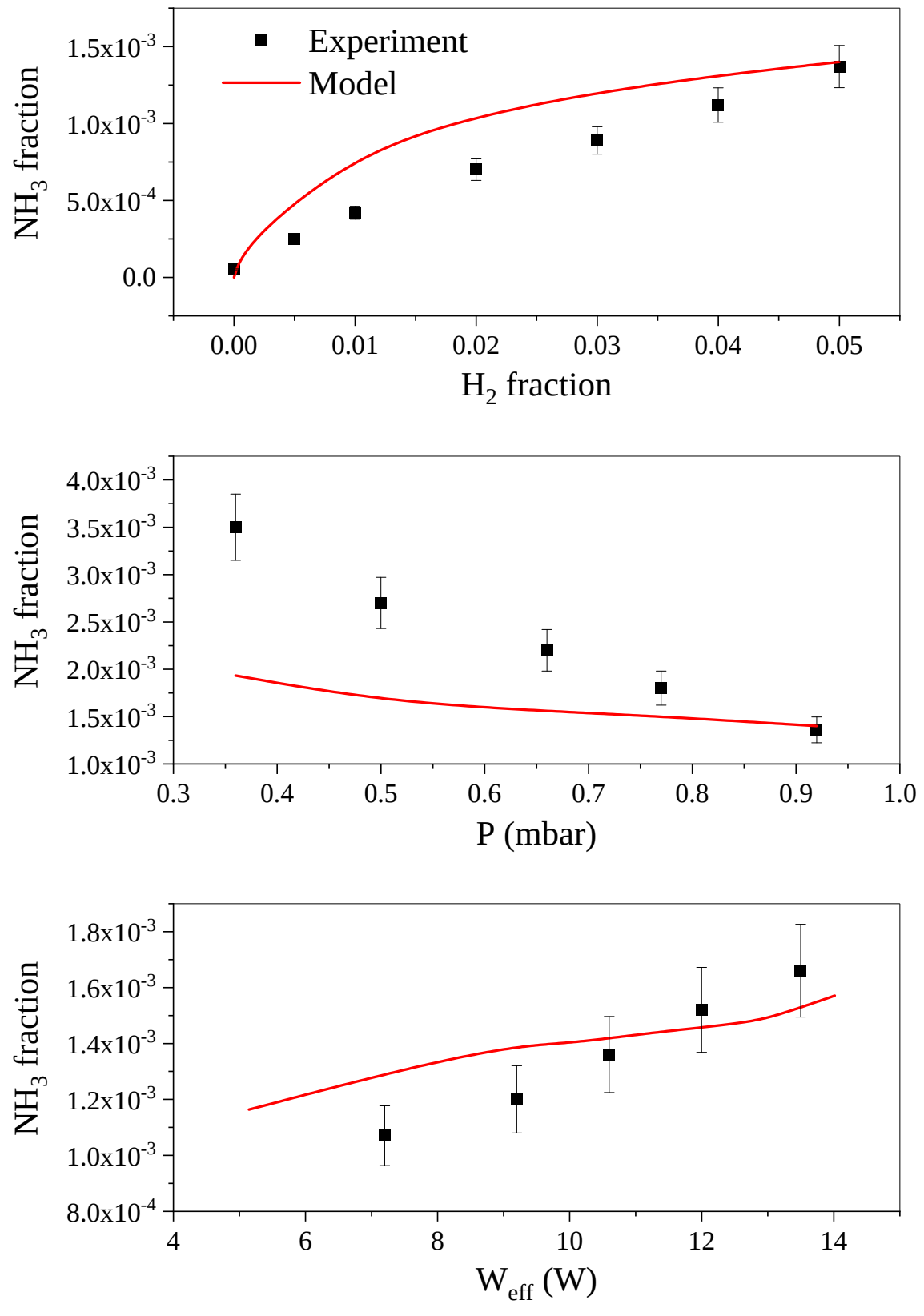

**Figure 10.** Simulations (curves) and measurements (points) of the $NH_3$ fraction, as a function of the $H_2$ fraction in the mixture (upper panel), pressure (middle) and power coupled (lower). The standard working conditions are 360 V voltage, 0.92 mbar pressure and 0.05 $H_2$ fraction; the results were obtained by fixing two of these parameters and varying the other.

As mentioned in section 2, the overall low abundances of $NH_3$ predicted by the model are the result of adopting a set of wall loss probabilities for low-reactivity surfaces, in reactions (w7)-w(18) (see Table A1; here, the most relevant parameters, for the sensitivity of model results, are those of reactions w9, w16 and w17). A comparison of model results using different sets of surface parameters can be found in Figure 11. Using the high-reactivity parameters of stainless steel [47] results in ammonia concentrations that are a factor of 3-4 times larger than the ones obtained for the current set of surface parameters, and thus much larger than the ones observed experimentally. Ammonia formation is indeed found to be relatively scarce in the discharge, and a possible explanation is the aluminium-alloy surface of the cage, leading to a much less efficient formation of $NH_3$ through the multi-step surface process.

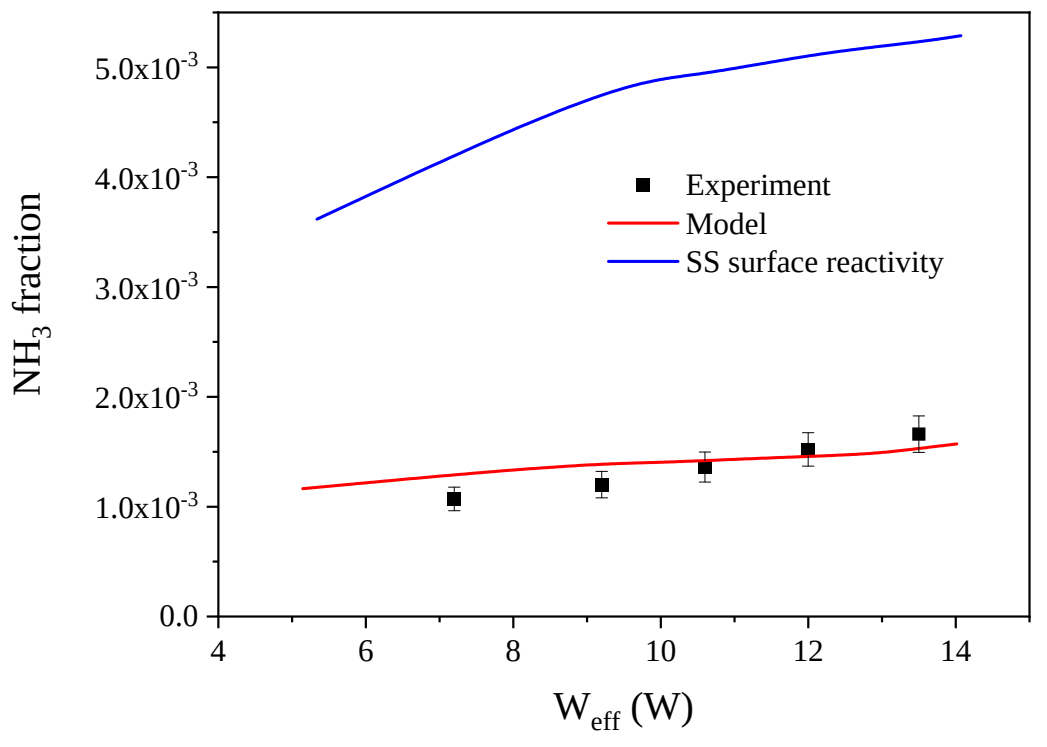

**Figure 11.** Simulations (curves) and measurements (points) of the $NH_3$ fraction, as a function of the power coupled, for the same conditions as in Figure 10. Simulations used the wall loss probabilities provided in Table A1 (red line) or the values for a stainless-steel surface proposed in [47] (blue line).

The main pathways for ammonia formation are shown in Figure 12, for the same discharge conditions as in the previous figure. It is apparent that the main formation mechanisms of $NH_3$ do not change in the range of conditions studied. These correspond to the final step of several Langmuir-Hinshelwood reactions (reaction w17 in Table A1), which accounts for about 80% of $NH_3$ formation at the wall, and the neutralization of the $NH_4^+$ ion at the reactor walls (w36). This conclusion is in agreement with other works studying the synthesis of $NH_3$ in $N_2$ + $H_2$ plasmas [49, 54]. Other processes, with minor influence in the formation of $NH_3$, do change in relevance, especially when increasing the $H_2$ fraction in the mixture (in this case the rate of Eley-Rideal reaction (w15) grows by two orders of magnitude), or the discharge power (in which case the ion neutralization by electron capture becoming as relevant as its recombination at the surfaces).

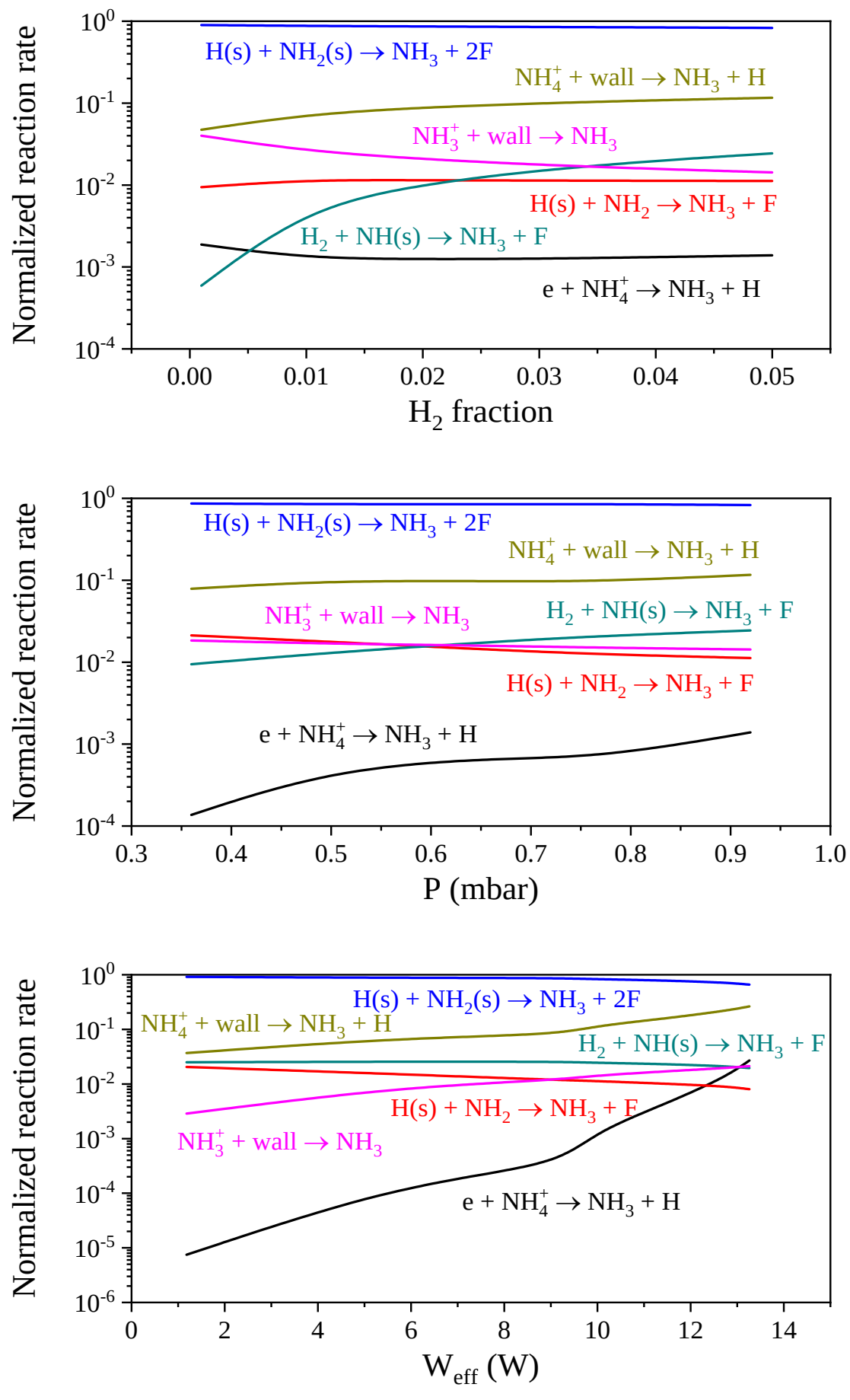

**Figure 12.** Normalized reaction rates for the main sources of ammonia, calculated for the same conditions as in Figure 10.

Surface coverage is thus the most important factor in the formation of $NH_3$. The density of adsorbed species is found to be only weakly dependent on most of the discharge parameters, and it is particularly impervious to the value of the secondary electron emission coefficient. It is, however, particularly sensitive to changes in the mixture composition. Figure 13 depicts the fraction of adsorbed species as a function of the $H_2$ fraction in the discharge. Notably, hydrogen is the most abundant adsorbed species, with only a remainder ~0.5% abundance in the gas phase. This is mainly due to the higher recombination coefficient of N at the surfaces (reaction w8, leading to the formation of $N_2$) compared to the one for $H_2$ (w7). For pure $N_2$ mixtures, where no formation of ammonia takes place, this efficient recombination of $N_2$ leads to a large amount of free surface sites (~20%). Note that the high H(s) abundance for mixtures with > 1% $H_2$ hinders the production of $NH_3$, as a higher N(s) abundance would make $NH_3$ formation more efficient [49]. Once NH(s) is formed at the surface, the subsequent reactions progress quickly, as evidenced by the low concentrations of both NH(s) and $NH_2$(s).

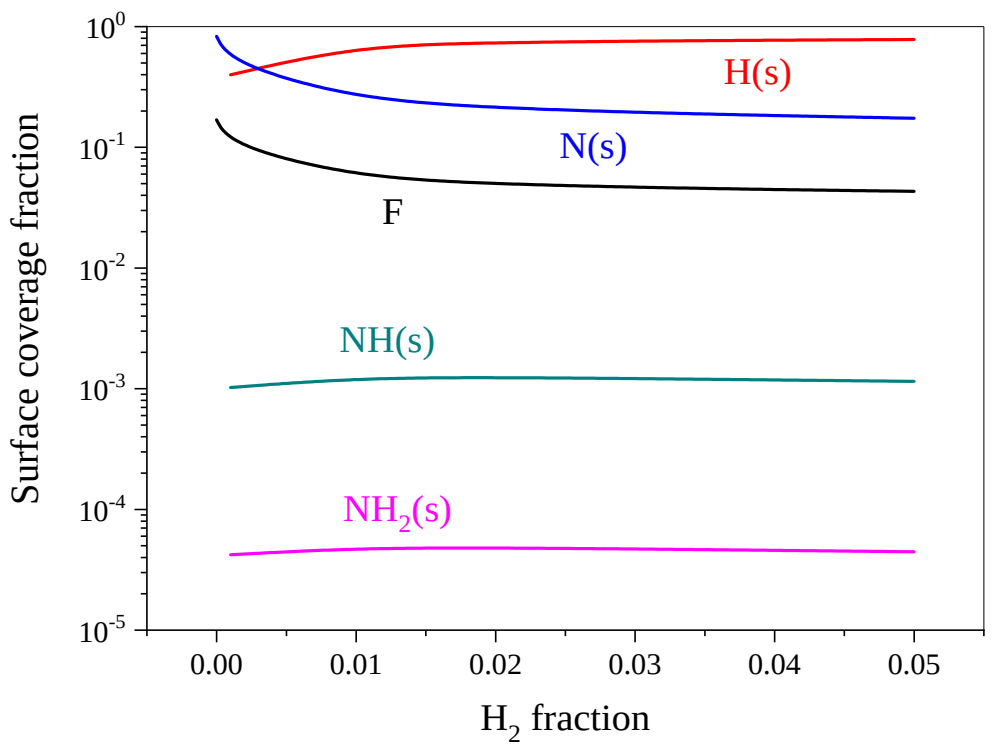

**Figure 13.** Calculated surface coverage fractions, as a function of the $H_2$ fraction in the mixture, for the same conditions as in Figure 10.

The densities of all neutral species in the model, displayed as a function of the $H_2$ fraction in the gas mixture, are plotted in the upper panel of Figure 14. The major neutrals in the plasma can be quickly identified in this figure, namely the precursor $N_2$ and $H_2$, the corresponding atoms, and $NH_3$. The hydrogen-containing species in this group readily grow when $H_2$ is added to the mixture, while $N_2$ and N remain mostly unchanged. The rest of the neutral species, all of the form $N_xH_y$, are found in very similar quantities about five orders of magnitude lower than that of $N_2$, and, in spite of their hydrogen content, are mostly unaffected by the increase in $H_2$ fraction.

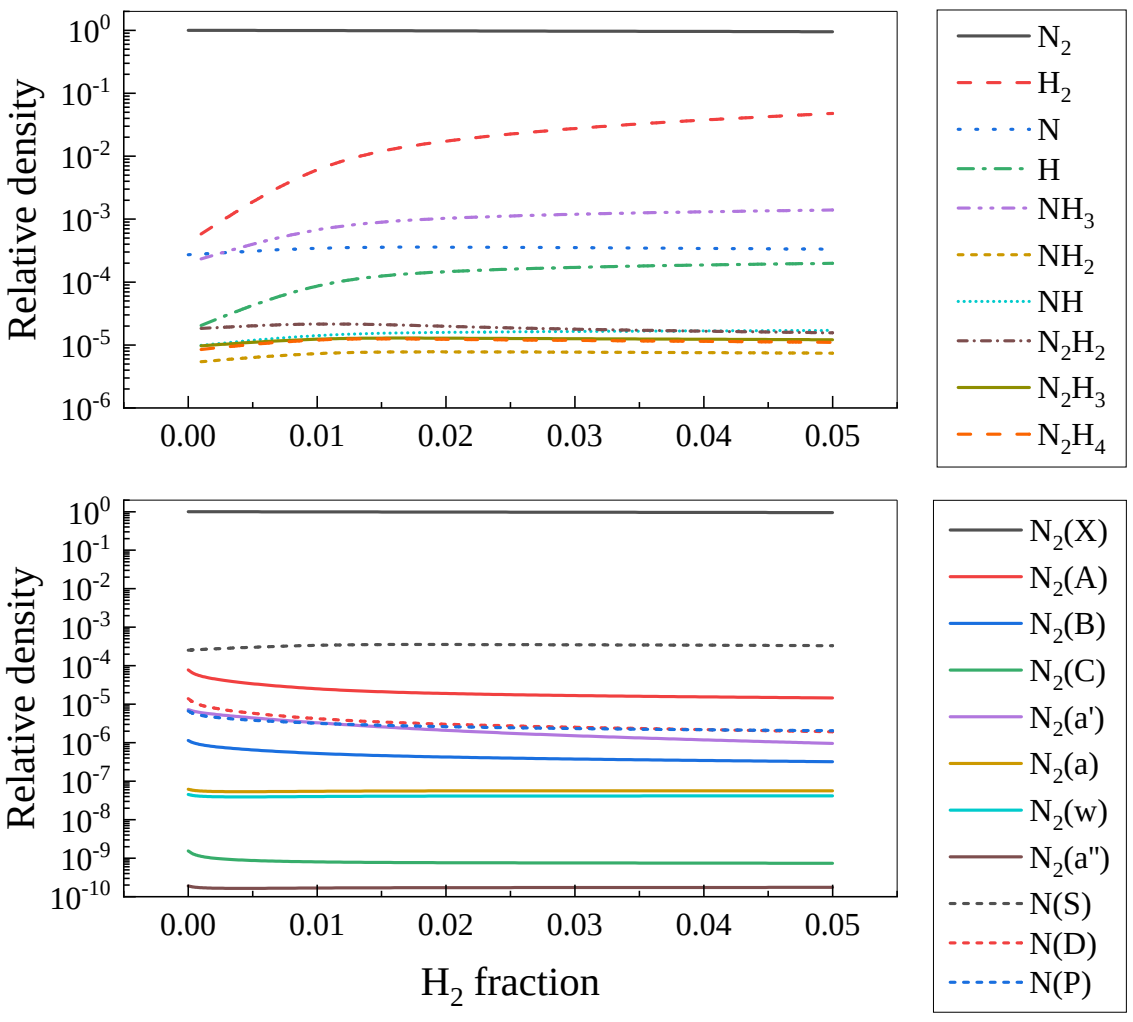

**Figure 14.** Relative densities obtained from the model as a function of the $H_2$ fraction for a discharge at 0.92 mbar and 10 W. Upper panel: neutral species. Lower panel: electronic states of molecular and atomic nitrogen.

The lower panel of Figure 14 shows the different electronic states considered in the model for $N_2$ and N. In the case of $N_2$, the ground state clearly dominates the distribution, with the density of the largest excited state, $N_2$(A), amounting to ~0.01% of that of $N_2$(X). Most of the densities of the excited states show a slight decrease as the $H_2$ fraction is increased. A somewhat similar situation is found for N atoms, although in this case the densities of the metastable states N(P) and N(D) are closer to the ground state (~1% of N(S)). The addition of $H_2$ also results in a very small increase of the N(S) density, while both excited states are found to decrease in a similar way to those of $N_2$.

Increasing the $H_2$ fraction in the mixture also leads to significant differences in the VDF of $N_2(X)$ and $H_2(X)$. This is shown in Figure 15. The VDF for $N_2(X)$ in pure $N_2$ plasmas shows a large plateau for states with $v > 10$, starting to decrease again only for the largest values of $v$ included in the model ($v > 40$). As soon as a small amount of $H_2$ is added, the length of this plateau is strongly reduced, with the second drop shifting to $v > 20$. The shape of the $N_2(X)$ VDF does not change significantly for mixtures with more than 1% $H_2$.

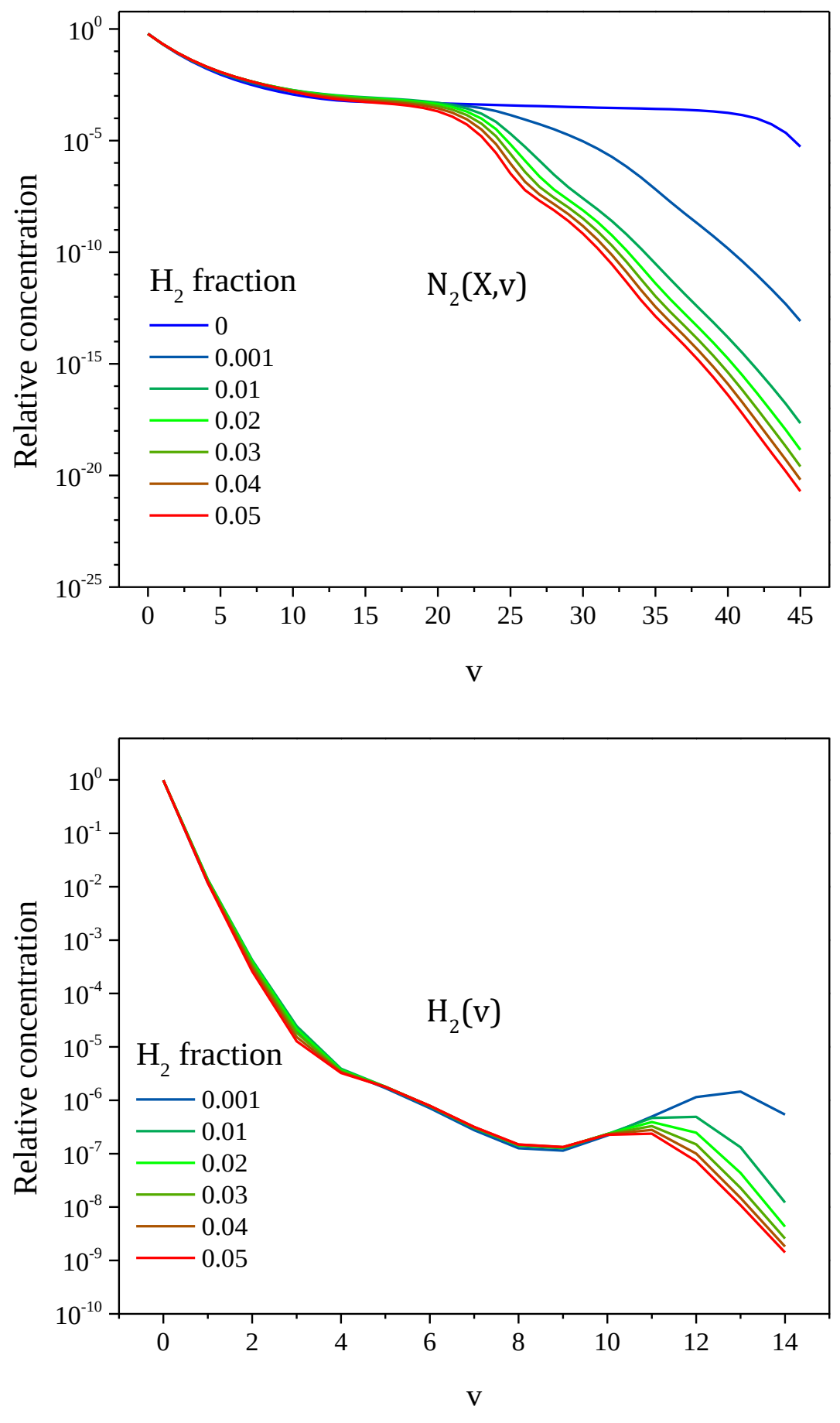

**Figure 15.** Calculated vibrational distribution functions for $N_2$ (upper panel) and $H_2$ (lower panel) molecules in the electronic ground-state, for various $H_2$ fractions in the mixture and the same conditions as in Figure 10.

In the case of $H_2$, the largest difference is also observed in the 0.1% - 1% $H_2$ fraction range. In this case, the $H_2(X)$ VDFs are virtually identical up to $v = 9$, for all mixtures. After this point, a population inversion is observed which is more significant when $H_2$ is less abundant in the mixture. This population inversion did not occur in the work of Gordiets et al. [52], although a small plateau for $v$-values between 9 and 12 was observed.

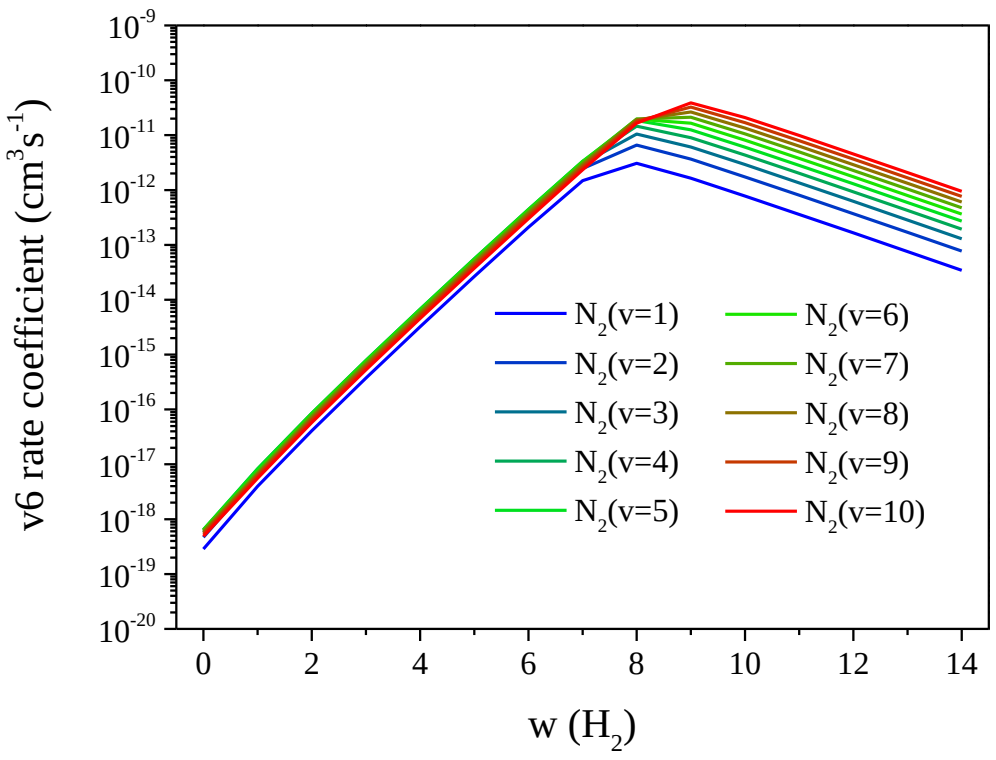

**Figure 16.** Rate coefficient for reaction v6, $N_2(X,v) + H_2(X,w) \rightarrow N_2(X,v{-}1) + H_2(X,w{+}1)$, as a function of the $H_2$ vibrational level.

In the present model, both the population inversion in the $H_2(X)$ VDF and the drop in the tail of the $N_2(X)$ VDF are due to the efficient V-V processes between both molecules (v8 and v9 in Table A1). As shown in Figure 16 for the monoquantum process, the V-V transfer with $N_2$ molecules in the lower vibrational states is most efficient for values of $v$ around 8, which leads to a depletion of these vibrational levels in $H_2$ and an increment in the population of the upper levels. A similar effect occurs for the higher vibrational levels of $N_2(X)$, which are more efficiently destroyed by collisions with $H_2$ than the lower ones.

## *3.3 Main processes in the plasma chemistry*

The purpose of this section is to identify the key mechanisms responsible for the creation and destruction of the major species present in the discharge. Tables 1, 2 and 3 show the normalized production and loss reaction rates of neutral, adsorbed and ionic species found in the plasma, respectively. They are calculated for a discharge at 0.92 mbar, 5% $H_2$ and 10 W. The relative rates depend on the specific conditions of the discharge and may vary depending on the concentration of the reactants. However, the main mechanisms listed here correspond to the most important production/loss kinetic processes of each species for most of the conditions sampled, as it was the case of the $NH_3$ production mechanisms discussed in the previous section. For brevity, only channels accounting for 5% or more of the total rate for each species are included in the tables.

**Table 1.** Main production and loss processes for the major neutral species of the discharge, calculated for a discharge with 5% $H_2$, 0.92 mbar and 10 W. Rates are given as a fraction of the total production/loss of each species. For the purposes of this table, $N_2(X)$ and $H_2$ mechanisms include those involving the vibrationally excited levels of these species.

| | **Production** | | | **Loss** | | |
|---|---|---|---|---|---|---|
| | | **Reaction** | **Rate** | | **Reaction** | **Rate** |
| **$N_2(X)$** | | Inflow $N_2$ | 0.88 | | Pumping | 0.88 |
| | n4 | $N_2(A) + N_2(X,5 \leq v \leq 14) \rightarrow N_2(B) + N_2(X,0)$ | 0.07 | n4 | $N_2(A) + N_2(X,5 \leq v \leq 14) \rightarrow N_2(B) + N_2(X,0)$ | 0.07 |
| **$H_2$** | | Inflow $H_2$ | 0.79 | | Pumping | 0.75 |
| | w7 | $H + H(s) \rightarrow H_2 + F$ | 0.20 | n15 | $N_2(a') + H_2(X) \rightarrow N_2(X) + 2H$ | 0.14 |
| **N(S)** | e20 | $e + N_2(X,v) \rightarrow e + N(S) + N(S)$ | 0.46 | w2 | $N + F \rightarrow N(s)$ | 0.27 |
| | e21 | $e + N_2(X,v) \rightarrow e + N(S) + N(D)$ | 0.25 | w8 | $N + N(s) \rightarrow N_2 + F$ | 0.21 |
| | n1 | $N_2(X,11 \leq v \leq 24) + N_2(X,11 \leq v \leq 24) \rightarrow N_2(X,0) + N(S) + N(S)$ | 0.06 | w9 | $N + H(s) \rightarrow NH(s)$ | 0.14 |
| | n21 | $N(P) + N_2 \rightarrow N(S) + N_2$ | 0.05 | n33 | $N(S) + NH_2 \rightarrow N_2(X) + 2H$ | 0.13 |
| | n44 | $H + NH \rightarrow N(S) + H_2(X)$ | 0.05 | n32 | $N(S) + NH \rightarrow N_2(X) + H$ | 0.13 |
| | n5 | $N_2(A) + N_2(X,14\leq v\leq 19) \rightarrow N_2(X,0) + N(S) + N(S)$ | 0.05 | n6 | $N_2(A) + N(S) \rightarrow N_2(X,6\leq v\leq 9) + N(P)$ | 0.09 |
| **H** | n15 | $N_2(a') + H_2(X) \rightarrow N_2(X) + 2H$ | 0.53 | w1 | $H + F \rightarrow H(s)$ | 0.57 |
| | n33 | $N(S) + NH_2 \rightarrow N_2(X) + 2H$ | 0.13 | w7 | $H + H(s) \rightarrow H_2 + F$ | 0.37 |
| | e22 | $e + H_2(X,v) \rightarrow e + 2H$ | 0.08 | | | |
| | n32 | $N(S) + NH \rightarrow N_2(X) + H$ | 0.06 | | | |
| | im6 | $N_2^+ + H_2 \rightarrow N_2H^+ + H$ | 0.05 | | | |
| | n35 | $N(D) + H_2(X) \rightarrow H + NH$ | 0.05 | | | |
| **$NH_3$** | w17 | $H(s) + NH_2(s) \rightarrow NH_3 + 2F$ | 0.83 | n36 | $N(D) + NH_3 \rightarrow NH_2 + NH$ | 0.51 |
| | w36 | $NH_4^+ + wall \rightarrow NH_3 + H$ | 0.12 | | Pumping | 0.31 |
| | | | | im24 | $N_2H^+ + NH_3 \rightarrow NH_4^+ + N_2$ | 0.12 |

As expected, the main sources and losses for both precursor molecules, $N_2$ and $H_2$, are related to the gas inflow into the reactor and the pumping (see Table 1). In both of these cases minor processes are also identified. In the case of $N_2$, it is a reaction with $N_2(A)$ that results in the conversion of vibrationally excited states to the ground state, resulting in no net change of the $N_2(X)$ density, but for $H_2$, the wall recycling of H atoms is shown to be a relevant process, while the dissociation through reaction with $N_2$(a') appears to be also an important mechanism.

Atoms are shown to have multiple significant mechanisms of creation and destruction. N(S) atoms are produced mostly by electron impact dissociation of $N_2$ and then lost to surface reactions, while, notably, H atoms are mostly created by the aforementioned reaction of $H_2$ with $N_2$(a'), with electron impact only accounting for ~8% of the total dissociation. They are then lost to surface reactions as well.

Finally, as was shown before, the main production channel of $NH_3$ is the Langmuir-Hinshelwood process involving H(s) and $NH_2$(S), while it is lost is not only to pumping but also through reaction with N(D) metastables.

**Table 2.** Same as Table 1, but for the adsorbed species at the reactor walls.

| | **Production** | | | **Loss** | | |
|---|---|---|---|---|---|---|
| | | **Reaction** | **Rate** | | **Reaction** | **Rate** |
| **H(s)** | w1 | $H + F \rightarrow H(s)$ | 0.87 | w7 | $H + H(s) \rightarrow H_2 + F$ | 0.56 |
| | w6 | $H_2 + 2F \rightarrow 2H(s)$ | 0.13 | w17 | $H(s) + NH_2(s) \rightarrow NH_3 + 2F$ | 0.17 |
| | | | | w16 | $NH(s) + H(s) \rightarrow NH_2(s) + F$ | 0.16 |
| | | | | w9 | $N + H(s) \rightarrow NH(s)$ | 0.11 |
| **N(s)** | w2 | $N + F \rightarrow N(s)$ | 0.99 | w8 | $N + N(s) \rightarrow N_2 + F$ | 0.78 |
| | | | | w10 | $H + N(s) \rightarrow NH(s)$ | 0.21 |
| **NH(s)** | w9 | $N + H(s) \rightarrow NH(s)$ | 0.66 | w16 | $NH(s) + H(s) \rightarrow NH_2(s) + F$ | 0.97 |
| | w10 | $H + N(s) \rightarrow NH(s)$ | 0.27 | | | |
| | w3 | $NH + F \rightarrow NH(s)$ | 0.07 | | | |
| **$NH_2$(s)** | w16 | $NH(s) + H(s) \rightarrow NH_2(s) + F$ | 0.94 | w17 | $H(s) + NH_2(s) \rightarrow NH_3 + 2F$ | 1.00 |

Surface chemistry, which is shown in Table 2, is somewhat simpler than the neutral chemistry previously analyzed. The hydrogen and nitrogen atoms adsorbed at the surface are mostly originating from atoms in the gas phase, and then a small number of reactions can happen from that point. The most favorable ones in our surface seems to be the recycling of the atoms into the precursors $N_2$ and $H_2$, however, NH can also be formed from both adsorbed atoms. The next two reactions are quite straightforward, with $NH_2$(s) being formed by surface association of H(s) and NH(s) and the final production of $NH_3$, which is the only possible loss mechanism for $NH_2$(s).

**Table 3.** Same as Table 1, but for the main ions in the plasma.

| | Production | | | Loss | | |
|---|---|---|---|---|---|---|
| | | Reaction | Rate | | Reaction | Rate |
| **$N_2^+$** | e28 | $e + N_2(X,0) \rightarrow N_2^+ + 2e$ | 1.00 | im6 | $N_2^+ + H_2 \rightarrow N_2H^+ + H$ | 0.85 |
| | | | | w26 | $N_2^+ + wall \rightarrow N_2$ | 0.12 |
| **$N_2H^+$** | im6 | $N_2^+ + H_2 \rightarrow N_2H^+ + H$ | 0.93 | w32 | $N_2H^+ + wall \rightarrow H + N_2$ | 0.75 |
| | im18 | $H_2^+ + N_2 \rightarrow N_2H^+ + H$ | 0.06 | im24 | $N_2H^+ + NH_3 \rightarrow NH_4^+ + N_2$ | 0.25 |
| **$NH_4^+$** | im24 | $N_2H^+ + NH_3 \rightarrow NH_4^+ + N_2$ | 0.96 | w36 | $NH_4^+ + wall \rightarrow NH_3 + H$ | 0.98 |
| **$NH_3^+$** | im7 | $N_2^+ + NH_3 \rightarrow NH_3^+ + N_2$ | 0.77 | w35 | $NH_3^+ + wall \rightarrow NH_3$ | 0.77 |
| | e43 | $e + NH_3 \rightarrow NH_3^+ + 2e$ | 0.20 | im32 | $NH_3^+ + NH_3 \rightarrow NH_2 + NH_4^+$ | 0.23 |
| **$H^-$** | im36 | $NH_2^- + H_2 \rightarrow NH_3 + H^-$ | 0.52 | ii5 | $H^- + NH_4^+ \rightarrow NH_3 + H_2$ | 0.56 |
| | e48 | $e + NH_3 \rightarrow NH_2 + H^-$ | 0.46 | im34 | $H^- + H \rightarrow H_2 + e$ | 0.43 |
| **$NH_2^-$** | e47 | $e + NH_3 \rightarrow NH_2^- + H$ | 1.00 | im36 | $NH_2^- + H_2 \rightarrow NH_3 + H^-$ | 0.98 |

The chemistry of the major positive ions in the discharge is also relatively simple and dominated by a handful of processes, as shown in Table 3. The main reactions accounting for a net production of charge are the ionization of $N_2$ and $NH_3$, which act as the start of a chain of hydrogenation reactions. $N_2^+$ is then lost through reaction with hydrogen to $N_2H^+$, which can then react with $NH_3$ to form $NH_4^+$. In the case of $NH_3^+$, it can directly react with $NH_3$ to also give $NH_4^+$. However, the $NH_4^+$ ion does not react with neutrals in the gas phase, instead being lost mainly to the reactor walls, which explains the large density of this ion found in the simulations. No other species is involved in this group of reactions besides those mentioned above, which might explain why this specific group of ions was so well reproduced in the comparison with the experiments of Figures 7 and 8.

Finally, the negative ions ($NH_2^-$, $H^-$) in the discharge also follow a simple chemistry. Both are formed by dissociative attachment of $NH_3$, but $NH_2^-$ can also be converted to $H^-$ through collisions with $H_2$, which explains why $H^-$ is the most abundant of the two.

## Conclusions

Capacitively coupled discharges in $N_2$-$H_2$ mixtures with hydrogen concentrations up to 5% have been studied by means of a hybrid code that couples a two-dimensional time-dependent fluid-type module, describing the transport of the charged particles, to a very complete zero-dimensional kinetic module for the nitrogen-hydrogen mixture. The results of the simulations have been compared to the experimental measurements described in paper I.

The electrical parameters obtained from the model are in good agreement with the measurements. To capture the behavior of the electron density with growing power, a module devoted to the simulation of the fast electrons generated by secondary emission must be included in the model. The inclusion of these fast electrons results in an exponential growth of the electron density at higher powers, which cannot be accounted for by the fluid code for slow electrons, and which can be fitted to the experimental values by tuning the secondary emission coefficient $\gamma$. A comparison between the slow and fast electrons in the model shows that, while both groups are responsible for ionization in the sheaths, fast electrons are present in the discharge bulk with high energies and produce ionizations in this region far more effectively than the slow electrons, which results in the observed increase in the electron density.

Ion fluxes to the reactor walls obtained from the simulations have been compared to the QMS measurements for charged particles. The ion distribution obtained experimentally is found to correspond to a high electron energy condition similar to the one obtained with the model for the sheath region. Further comparison of the ion fluxes in this region with the experiments shows a good agreement in the fluxes of the major positive ions in the discharge ($N_2^+$, $N_2H^+$, $NH_4^+$), which also follow the experimental trends when varying the $H_2$ fraction, the pressure and the coupled power. However, the behavior of the minor ions in the discharge showed some discrepancies between model and experiment. Ion fluxes are also shown not to be representative of the actual ion composition of the discharge, due to the different distribution of ions in the sheath and in the bulk.

Ion and neutral chemistry are found to be heavily influenced by the formation of $NH_3$ at the reactor walls. The comparison between the predicted $NH_3$ fraction and the experimental one implies that the surface reactivity of the chamber must be significantly low compared to the one for stainless steel. With this lower reactivity, the trends of the $NH_3$ fraction as a function of $H_2$, pressure, and power are correctly reproduced, although the quantitative agreement could be improved. The analysis of the different formation mechanisms of $NH_3$ evinces that surface reactions are largely responsible for its formation, and only at large powers is ion neutralization a somewhat competing mechanism.

A significant effect of hydrogen addition is also observed in the vibrational distributions of $N_2$ and $H_2$. The populations of the higher vibrational levels are found to be especially sensitive to collisions between the two species, resulting in a decrease of the VDF tail with growing $H_2$.

The main production and loss processes for the major species in the plasma were extracted from the model. Neutral chemistry was found to be more complex than heterogeneous and ionic chemistry, where only a few reactions were responsible for the final distributions obtained from the simulations.

## Acknowledgements

L. Marques and M. Jiménez-Redondo acknowledge the financial support of the Portuguese Foundation for Science and Technology (FCT) in the framework of the Strategic Funding UIDB/04650/2019 and project UTAP-EXPL/NTec/0107/2017.

L.L. Alves acknowledges the financial support of the Portuguese Foundation for Science and Technology (FCT) through project UID/FIS/50010/2019 and grant SFRH/BSAB/150267/2019.

N. Carrasco acknowledges the financial support of the European Research Council (ERC Starting Grant PRIMCHEM, Grant agreement no. 636829).

A. Chatain acknowledges ENS Paris-Saclay Doctoral Program.

# Appendix

**Table A1.** Complete set of reactions included in the model. Rate coefficients are given in $cm^3\ s^{-1}$, except for three body reactions ($cm^6\ s^{-1}$) and emission rates ($s^{-1}$). Sticking and recombination coefficients are non-dimensional; $E_{LH}$ is given in eV; $T_g$ is given in K. The label *f(E/N)* means that the rate coefficient was obtained by integration of the corresponding cross section over the EEDF, self-consistently calculated by the model.

| Vibrational processes | | | |
|---|---|---|---|
| v1 | e–V | $e + N_2(X,v) \leftrightarrow e + N_2(X,w)$ ; $v$=0–9 ; $w$=($v$+1)–10 | [66, 71] |
| v2 | V–T | $N_2(X,v) + N_2 \rightarrow N_2(X,w) + N_2$ | [60] |
| v3 | V–V | $N_2(X,v) + N_2(X,w) \rightarrow N_2(X,v+1) + N_2(X,w-1)$ | [60] |
| v4 | | $N_2(X,v)$ + wall $\rightarrow N_2(X,v-1)$ | [60] |
| v5 | V–T | $N_2(X,v) + H_2 \rightarrow N_2(X,w) + H_2$ | [52] |
| v6 | V–T | $N_2(X,v) + N \rightarrow N_2(X,w) + N$ | [52] |
| v7 | V–T | $N_2(X,v) + H \rightarrow N_2(X,w) + H$ | [52] |
| v8 | V–V | $N_2(X,v) + H_2(X,w) \leftrightarrow N_2(X,v+1) + H_2(X,w-1)$ | [52] |
| v9 | V–V | $N_2(X,v) + H_2(X,w) \leftrightarrow N_2(X,v+2) + H_2(X,w-1)$ | [52] |
| v10 | e–V | $e + H_2(X,v) \leftrightarrow e + H_2(X,v+i)$ ; $v$=0–13 ; $i$=1–3 | [66] |
| v11 | V–T | $H_2(X,v) + H_2 \rightarrow H_2(X,w) + H_2$ | [61] |
| v12 | V–T | $H_2(X,v) + H \rightarrow H_2(X,w) + H$ | [52] |
| v13 | V–V | $H_2(X,v) + H_2(X,w) \leftrightarrow H_2(X,v+1) + H_2(X,w-1)$ | [61] |
| v14 | | $H_2(X,v)$ + wall $\rightarrow H_2(X,v-1)$ | [61] |
| v15 | E-V | $e + H_2(X,v) \rightarrow e + H_2(B,C) \rightarrow e + H_2(X,w)$ | [61, 66] |

| Electron impact processes | | | |
|---|---|---|---|
| Excitation and deexcitation | | Rate coefficient | Ref. |
| e1 | $e + N_2(X,0) \rightarrow e + N_2(A)$ | $f(E/N)$ | [66] |
| e2 | $e + N_2(X,0) \rightarrow e + N_2(B)$ | $f(E/N)$ | [66] |
| e4 | $e + N_2(X,0) \rightarrow e + N_2(B')$ | $f(E/N)$ | [66] |
| e3 | $e + N_2(X,0) \rightarrow e + N_2(C)$ | $f(E/N)$ | [66] |
| e5 | $e + N_2(X,0) \rightarrow e + N_2(a')$ | $f(E/N)$ | [66] |
| e6 | e + $N_2$(X,0) → e + $N_2$(a) | *f*(*E*/*N*) | [66] |
| e7 | e + $N_2$(X,0) → e + $N_2$(w) | *f*(*E*/*N*) | [66] |
| e8 | e + $N_2$(X,0) → e + $N_2$(a'') | *f*(*E*/*N*) | [66] |
| e9 | e + $N_2$(A) → e + $N_2$(X,0) | *f*(*E*/*N*) | [66] |
| e10 | e + $N_2$(A) → e + $N_2$(B) | *f*(*E*/*N*) | [66] |
| e11 | e + $N_2$(A) → e + $N_2$(C) | *f*(*E*/*N*) | [66] |
| e12 | e + $N_2$(B) → e + $N_2$(A) | *f*(*E*/*N*) | [66] |
| e13 | e + $N_2$(C) → e + $N_2$(A) | *f*(*E*/*N*) | [66] |
| e14 | e + N(S) → e + N(D) | *f*(*E*/*N*) | [66] |
| e15 | e + N(S) → e + N(P) | *f*(*E*/*N*) | [66] |
| e16 | e + N(D) → e + N(S) | *f*(*E*/*N*) | [66] |
| e17 | e + N(D) → e + N(P) | *f*(*E*/*N*) | [66] |
| e18 | e + N(P) → e + N(S) | *f*(*E*/*N*) | [66] |
| e19 | e + N(P) → e + N(D) | *f*(*E*/*N*) | [66] |
| Dissociation | | Rate coefficient | |
| e20 | e + $N_2$(X,*v*) → e + N(S) + N(S) | *f*(*E*/*N*) | [60] |
| e21 | e + $N_2$(X,*v*) → e + N(S) + N(D) | *f*(*E*/*N*) | [60] |
| e22 | e + $H_2$(X,*v*) → e + 2H | *f*(*E*/*N*) | [66] |
| e23 | e + NH → N + H + e | *f*(*E*/*N*) | [65, 83] |
| e24 | e + $NH_2$ → N + $H_2$ + e | *f*(*E*/*N*) | [65, 83] |
| e25 | e + $NH_2$ → NH + H + e | *f*(*E*/*N*) | [65, 83] |
| e26 | e + $NH_3$ → $NH_2$ + H + e | *f*(*E*/*N*) | [83] |
| e27 | e + $NH_3$ → NH + $H_2$ + e | *f*(*E*/*N*) | [83] |
| Ionization | | Rate coefficient | |
| e28 | e + $N_2$(X,0) → $N_2^+$ + 2e | *f*(*E*/*N*) | [66] |
| e29 | e + $N_2$(X,0) → $N^+$ + N + 2e | *f*(*E*/*N*) | [66, 90] |
| e30 | e + $N_2$(A) → $N_2^+$ + 2e | *f*(*E*/*N*) | [66] |
| e31 | e + $N_2$(B) → $N_2^+$ + 2e | *f*(*E*/*N*) | [66] |
| e32 | e + $N_2$(a') → $N_2^+$ + 2e | *f*(*E*/*N*) | [66] |
| e33 | e + $N_2$(a) → $N_2^+$ + 2e | *f*(*E*/*N*) | [66] |
| e34 | e + $N_2$(w) → $N_2^+$ + 2e | *f*(*E*/*N*) | [66] |
| e35 | e + $H_2$ → $H_2^+$ + 2e | *f*(*E*/*N*) | [66, 91] |
| e36 | e + $H_2$ → $H^+$ + H + 2e | *f*(*E*/*N*) | [66, 91] |
| e37 | e + N → $N^+$ + 2e | *f*(*E*/*N*) | [66] |
| e38 | e + H → $H^+$ + 2e | *f*(*E*/*N*) | [66] |
| e39 | e + NH → $NH^+$ + 2e | *f*(*E*/*N*) | [92] |
| e40 | e + NH → $N^+$ + H + 2e | *f*(*E*/*N*) | [92] |
| e41 | e + $NH_2$ → $NH_2^+$ + 2e | *f*(*E*/*N*) | [92] |
| e42 | e + $NH_2$ → $NH^+$ + H + 2e | *f*(*E*/*N*) | [92] |
| e43 | $e + NH_3 \rightarrow NH_3^+ + 2e$ | $f(E/N)$ | [83] |
| e44 | $e + NH_3 \rightarrow NH_2^+ + H + 2e$ | $f(E/N)$ | [83, 93] |
| e45 | $e + NH_3 \rightarrow NH^+ + H_2 + 2e$ | $f(E/N)$ | [83, 93] |
| e46 | $e + NH_3 \rightarrow N^+ + H_2 + H + 2e$ | $f(E/N)$ | [83, 93] |
| e47 | $e + NH_3 \rightarrow H^+ + NH_2 + 2e$ | $f(E/N)$ | [83, 93] |
| Attachment and detachment | | Rate coefficient | |
| e45 | $e + H_2 \rightarrow H + H^-$ | $f(E/N)$ | [94] |
| e46 | $e + NH_2 \rightarrow NH + H^-$ | $f(E/N)$ | [83, 95] |
| e47 | $e + NH_3 \rightarrow NH_2^- + H$ | $f(E/N)$ | [83, 95] |
| e48 | $e + NH_3 \rightarrow NH_2 + H^-$ | $f(E/N)$ | [83, 95] |
| e49 | $e + H^- \rightarrow H + 2e$ | $f(E/N)$ | [96] |
| Neutralization | | Rate coefficient | |
| e50 | $e + N^+ \rightarrow N$ | $f(E/N)$ | [65] |
| e51 | $e + N_2^+ \rightarrow N + N$ | $f(E/N)$ | [60] |
| e52 | $e + N_3^+ \rightarrow N_2 + N$ | $f(E/N)$ | [71] |
| e53 | $e + N_4^+ \rightarrow N_2 + N_2$ | $f(E/N)$ | [60] |
| e54 | $e + H^+ \rightarrow H$ | $f(E/N)$ | [97] |
| e55 | $e + H_2^+ \rightarrow H + H$ | $f(E/N)$ | [97] |
| e56 | $e + H_3^+ \rightarrow H_2 + H$ | $f(E/N)$ | [97] |
| e57 | $e + N_2H^+ \rightarrow N_2 + H$ | $f(E/N)$ | [49] |
| e58 | $e + NH^+ \rightarrow N + H$ | $f(E/N)$ | [49] |
| e59 | $e + NH_2^+ \rightarrow N + 2H$ | $f(E/N)$ | [49] |
| e60 | $e + NH_2^+ \rightarrow NH + H$ | $f(E/N)$ | [49] |
| e61 | $e + NH_3^+ \rightarrow NH + 2H$ | $f(E/N)$ | [49] |
| e62 | $e + NH_3^+ \rightarrow NH_2 + H$ | $f(E/N)$ | [49] |
| e63 | $e + NH_4^+ \rightarrow NH_2 + 2H$ | $f(E/N)$ | [49] |
| e64 | $e + NH_4^+ \rightarrow NH_3 + H$ | $f(E/N)$ | [49] |
| Neutral-neutral processes | | | |
| Collisional excitation and deexcitation | | Rate coefficient | |
| n1 | $N_2(X,11\leq v\leq 24) + N_2(X,11\leq v\leq 24) \rightarrow N_2(X,0) + N(S) + N(S)$ | $3.50\times10^{-15}$ | [60] |
| n2 | $N_2(A) + N_2(A) \rightarrow N_2(B) + N_2(X,0)$ | $7.7\times10^{-11}$ | [60] |
| n3 | $N_2(A) + N_2(A) \rightarrow N_2(C) + N_2(X,0)$ | $1.50\times10^{-10}$ | [60] |
| n4 | $N_2(A) + N_2(X,5\leq v\leq 14) \rightarrow N_2(B) + N_2(X,0)$ | $2.0\times10^{-11}$ | [60] |
| n5 | $N_2(A) + N_2(X,14\leq v\leq 19) \rightarrow N_2(X,0) + N(S) + N(S)$ | $1.5\times10^{-12}$ | [60] |
| n6 | $N_2(A) + N(S) \rightarrow N_2(X,6\leq v\leq 9) + N(P)$ | $4.0\times10^{-11}$ | [60] |
| n7 | $N_2(A) + H \rightarrow N_2(X,0) + H$ | $5.0\times10^{-11}$ | [52] |
| n8 | $N_2(A) + H_2 \rightarrow N_2(X) + 2H$ | $2.0\times10^{-10}\times\exp(-3500/T_g)$ | [52] |
| n9 | $N_2(A) + NH_3 \rightarrow N_2(X,0) + NH_3$ | $1.6\times10^{-10}$ | [52] |
| n10 | $N_2(B) + N_2 \rightarrow N_2(A) + N_2$ | $0.95\times3.0\times10^{-11}$ | [60] |
| n11 | $N_2(B) + N_2 \rightarrow N_2(X,0) + N_2$ | $0.05\times3.0\times10^{-11}$ | [60] |
| n12 | $N_2(B) + H_2 \rightarrow N_2(A) + H_2$ | $2.5\times10^{-11}$ | [52] |
| n13 | $N_2(a') + N_2 \rightarrow N_2(B) + N_2$ | $1.90\times10^{-13}$ | [60] |
| n14 | $N_2(a') + H \rightarrow N_2(X,0) + H$ | $1.5\times10^{-11}$ | [52] |
| n15 | $N_2(a') + H_2(X) \rightarrow N_2(X) + 2H$ | $2.6\times10^{-11}$ | [52] |
| n16 | $N_2(a) + N_2 \rightarrow N_2(a') + N_2$ | $2.0\times10^{-11}$ | [60] |
| n17 | $N_2(w) + N_2 \rightarrow N_2(a) + N_2$ | $1.0\times10^{-11}$ | [60] |
| n18 | $N_2(a'') + N_2 \rightarrow$ products | $2.3\times10^{-10}$ | [60] |
| n19 | $N(S) + N(S) + N_2 \rightarrow N_2(B) + N_2$ | $8.27\times10^{-34} \exp(500/T_g)$ | [60] |
| n20 | $N(D) + N_2 \rightarrow N(S) + N_2$ | $1.0\times10^{-13} \exp(-510/T_g)$ | [60] |
| n21 | $N(P) + N_2 \rightarrow N(S) + N_2$ | $6.0\times10^{-14}$ | [60] |
| n22 | $N(P) + N_2(X,v\geq10) \rightarrow N(S) + N_2(A)$ | $1.0\times10^{-10} \exp(-1300/T_g)$ | [60] |
| n23 | $N(P) + N(S) \rightarrow N(S) + N(S)$ | $1.2\times10^{-12}$ | [60] |
| n24 | $N(P) + N(S) \rightarrow N(S) + N(D)$ | $6.0\times10^{-13}$ | [60] |
| Penning ionization | | Rate coefficient | |
| n26 | $N_2(A) + N_2(a') \rightarrow N_4^+ + e$ | $0.5\times1.0\times10^{-11}$ | [60] |
| n27 | $N_2(A) + N_2(a') \rightarrow N_2(X,0) + N_2^+ + e$ | $0.5\times1.0\times10^{-11}$ | [60] |
| n28 | $N_2(a') + N_2(a') \rightarrow N_4^+ + e$ | $0.5\times5.0\times10^{-11}$ | [60] |
| n29 | $N_2(a') + N_2(a') \rightarrow N_2(X,0) + N_2^+ + e$ | $0.5\times5.0\times10^{-11}$ | [60] |
| n30 | $N(D) + N(P) \rightarrow N_2^+ + e$ | $1.0\times10^{-13}$ | [60] |
| n31 | $N_2(a') + N(P) \rightarrow N_3^+ + e$ | $10^{-11}$ | [71] |
| Neutral-neutral reactions | | Rate coefficient | |
| n32 | $N(S) + NH \rightarrow N_2(X) + H$ | $5.0\times10^{-11}$ | [52] |
| n33 | $N(S) + NH_2 \rightarrow N_2(X) + 2H$ | $1.2\times10^{-10}$ | [52] |
| n34 | $N(S) + H_2(X,v) \rightarrow H + NH$ | $f(v,T_g)$ | [52] |
| n35 | $N(D) + H_2(X) \rightarrow H + NH$ | $2.3\times10^{-12}$ | [52] |
| n36 | $N(D) + NH_3 \rightarrow NH_2 + NH$ | $1.1\times10^{-10}$ | [52] |
| n37 | $N(P) + H_2(X) \rightarrow H + NH$ | $2.5\times10^{-14}$ | [52] |
| n38 | $N + N + M \rightarrow N_2 + M$ | $8.3\times10^{-34} \exp(500/T_g)$ | [52] |
| n39 | $H_2 + NH_2 \rightarrow NH_3 + H$ | $5.4\times10^{-11} \exp(-6492/T_g)$ | [52] |
| n40 | $H_2 + N + M \rightarrow NH_2 + M$ | $1.0\times10^{-36}$ | [52] |
| n41 | $H_2 + NH + M \rightarrow NH_3 + M$ | $2.5\times10^{-35} (T_g/300)^{2.63} \exp(1700/T_g)$ | [52] |
| n42 | $H + H + M \rightarrow H_2 + M$ | $8.3\times10^{-33} (300/T_g)$ | [52] |
| n43 | $H + N + M \rightarrow NH + M$ | $5.0\times10^{-32}$ | [65] |
| n44 | $H + NH \rightarrow N(S) + H_2(X)$ | $5.4\times10^{-11} \exp(-165/T_g)$ | [52] |
| n45 | $H + NH_2 \rightarrow H_2(X) + NH$ | $6.6\times10^{-11} \exp(-1840/T_g)$ | [52] |
| n46 | $H + NH_2 + M \rightarrow NH_3 + M$ | $5.5\times10^{-30}$ | [52] |
| n47 | $H + NH_3 \rightarrow H_2(X) + NH_2$ | $8.4\times10^{-14} \exp(-4760/T_g) (T_g/300)^{4.1}$ | [52] |
| n48 | $NH + NH \rightarrow N_2(X) + H_2(X)$ | $5.0\times10^{-14} (T_g/300)^{0.5}$ | [52] |
| n49 | $NH + NH \rightarrow N(S) + NH_2$ | $1.7\times10^{-12} (T_g/300)^{1.5}$ | [52] |
| n50 | $NH + NH \rightarrow 2H + N_2(X)$ | $8.5\times10^{-11}$ | [52] |
| n51 | $NH + NH \rightarrow N_2H_2$ | $3.5\times10^{-12}$ | [65] |
| n52 | $NH + NH_2 \rightarrow N(S) + NH_3$ | $1.66\times10^{-12}$ | [52] |
| n53 | $NH + NH_2 \rightarrow H + N_2H_2$ | $2.5\times10^{-9} (300/T_g)^{0.5}$ | [65] |
| n54 | $NH + NH_2 \rightarrow N_2H_3$ | $1.2\times10^{-10}$ | [65] |
| n55 | $NH + NH_3 + M \rightarrow N_2H_4 + M$ | $4.0\times10^{-35}$ | [65] |
| n56 | $NH + NH_3 + NH_3 \rightarrow N_2H_4 + NH_3$ | $1.0\times10^{-33}$ | [65] |
| n57 | $NH_2 + NH_2 \rightarrow N_2H_2 + H_2$ | $1.3\times10^{-12}$ | [65] |
| n58 | $NH_2 + NH_2 \rightarrow NH_3 + NH$ | $1.8\times10^{-14} (T_g/300)^{2.79} \exp(-660/T_g)$ | [65] |
| n59 | $NH_2 + NH_2 + NH_3 \rightarrow N_2H_4 + NH_3$ | $6.9\times10^{-30}$ | [65] |
| n60 | $N_2H_2 + H \rightarrow N_2 + H_2 + H$ | $4.5\times10^{-13} (T_g/300)^{2.63} \exp(115/T_g)$ | [65] |
| n61 | $N_2H_2 + NH_2 \rightarrow N_2 + H + NH_3$ | $1.5\times10^{-13} (T_g/300)^{4.05} \exp(810/T_g)$ | [65] |
| n62 | $N_2H_3 + H \rightarrow NH_2 + NH_2$ | $2.7\times10^{-12}$ | [65] |
| n63 | $N_2H_3 + N_2H_3 \rightarrow NH_3 + NH_3 + N_2$ | $5.0\times10^{-12}$ | [65] |
| n64 | $N_2H_3 + N_2H_3 \rightarrow N_2H_4 + N_2H_2$ | $2.0\times10^{-11}$ | [65] |
| n65 | $N_2H_4 + H \rightarrow N_2H_3 + H_2$ | $1.2\times10^{-11} \exp(-1260/T_g)$ | [65] |
| n66 | $N_2H_4 + N \rightarrow N_2H_2 + NH_2$ | $1.3\times10^{-13}$ | [65] |
| n67 | $N_2H_4 + NH_2 \rightarrow N_2H_3 + NH_3$ | $5.2\times10^{-13}$ | [65] |

| Ion-molecule reactions | | Rate coefficient | |
|---|---|---|---|
| im1 | $N^+ + H \rightarrow N + H^+$ | $2.0\times10^{-9}$ | [65] |
| im2 | $N^+ + H_2 \rightarrow H + NH^+$ | $5.6\times10^{-10}$ | [65] |
| im3 | $N^+ + M \rightarrow N + M^+$   M=$NH_2$, $NH_3$ | $2.4\times10^{-9}$ | [65] |
| im4 | $N^+ + N_2 + N_2 \rightarrow N_3^+ + N_2$ | $1.7\times10^{-29} \exp(300/T_g)^{2.1}$ | [71] |
| im5 | $N_2^+ + N_2 + N_2 \rightarrow N_4^+ + N_2$ | $6.8\times10^{-29} (300/T_g)^{1.64}$ | [52] |
| im6 | $N_2^+ + H_2 \rightarrow N_2H^+ + H$ | $1.7\times10^{-9}$ | [52] |
| im7 | $N_2^+ + NH_3 \rightarrow NH_3^+ + N_2$ | $1.95\times10^{-9}$ | [49] |
| im8 | $N_2^+ + N(S) + N_2 \rightarrow N_3^+ + N_2$ | $0.9\times10^{-29} \exp(400/T_g)$ | [71] |
| im9 | $N_2^+ + N_2(A) \rightarrow N_3^+ + N$ | $3\times10^{-10}$ | [71] |
| im10 | $N_3^+ + N(S) \rightarrow N_2^+ + N_2$ | $6.6\times10^{-11}$ | [71] |
| im11 | $N_4^+ + N_2 \rightarrow N_2^+ + N_2(X,0) + N_2$ | $2.1\times10^{-16} \exp(T_g/120)$ | [52] |
| im12 | $N_4^+ + H_2 \rightarrow H + N_2 + N_2H^+$ | $1.139\times10^{-8}$ | [52] |
| im13 | $N_4^+ + H_2 \rightarrow 2N_2 + H_2^+$ | $3.0\times10^{-10} \exp(-1800/T_g)$ | [52] |
| im14 | $H^+ + H_2 \rightarrow H + H_2^+$ | $2.5\times10^{-9}$ | [98] |
| im15 | $H^+ + 2H_2 \rightarrow 2H + H_3^+$ | $3.1\times10^{-29} (300/T_g)^{0.5}$ | [98] |
| im16 | $H^+ + M \rightarrow H + M^+$   M=$NH_x$   x=1–3 | $5.0\times10^{-11}$ | [65] |
| im17 | $H_2^+ + H_2 \rightarrow H_3^+ + H$ | $2.1\times10^{-9}$ | [52] |
| im18 | $H_2^+ + N_2 \rightarrow N_2H^+ + H$ | $1.95\times10^{-9}$ | [52] |
| im19 | $H_2^+ + H \rightarrow H_2 + H^+$ | $6.4\times10^{-10}$ | [65] |
| im20 | $H_2^+ + NH_x \rightarrow H + NH_{x+1}^+$ x=1–3 | $5.0\times10^{-11}$ | [65] |
| im21 | $H_2^+ + M \rightarrow H_2 + M^+$ | $5.0\times10^{-10}$ | [65] |
| im22 | $H_3^+ + N_2 \rightarrow N_2H^+ + H_2$ | $1.8\times10^{-9}$ | [52] |
| im23 | $H_3^+ + NH_3 \rightarrow H_2 + NH_4^+$ | $4.4\times10^{-9}$ | [65] |
| im24 | $N_2H^+ + NH_3 \rightarrow NH_4^+ + N_2$ | $2.3\times10^{-9}$ | [49] |
| im25 | $NH^+ + H_2 \rightarrow H + NH_2^+$ | $1.0\times10^{-9}$ | [65] |
| im26 | $NH^+ + NH_2 \rightarrow NH + NH_2^+$ | $1.8\times10^{-9}$ | [65] |
| im27 | $NH^+ + NH_3 \rightarrow NH + NH_3^+$ | $1.8\times10^{-9}$ | [65] |
| im28 | $NH^+ + NH_3 \rightarrow N + NH_4^+$ | $6.0\times10^{-10}$ | [65] |
| im29 | $NH_2^+ + NH_3 \rightarrow NH_2 + NH_3^+$ | $1.1\times10^{-9}$ | [65] |
| im30 | $NH_2^+ + NH_3 \rightarrow NH + NH_4^+$ | $1.1\times10^{-9}$ | [65] |
| im31 | $NH_3^+ + NH_3 \rightarrow H + NH_2 + NH_3^+$ | $2.0\times10^{-10}$ | [65] |
| im32 | $NH_3^+ + NH_3 \rightarrow NH_2 + NH_4^+$ | $2.2\times10^{-9}$ | [65] |
| im33 | $NH_3^+ + H_2 \rightarrow H + NH_4^+$ | $4.0\times10^{-13}$ | [65] |
| im34 | $H^- + H \rightarrow H_2 + e$ | $1.8\times10^{-9}$ | [52] |
| im35 | $H^- + NH_3 \rightarrow H_2 + NH_2^-$ | $8.8\times10^{-13}$ | [65] |
| im36 | $NH_2^- + H_2 \rightarrow NH_3 + H^-$ | $2.3\times10^{-11}$ | [65] |

| Ion-ion recombination | | Rate coefficient | |
|---|---|---|---|
| ii1 | $H^- + H^+ \rightarrow 2H$ | $8.73\times10^{-9}\ (300/T_g)^{0.5}$ | [99] |
| ii2 | $H^- + H_2^+ \rightarrow H_2 + H$ | $2.91\times10^{-9}\ (300/T_g)^{0.5}$ | [99] |
| ii3 | $H^- + H_3^+ \rightarrow 2H_2$ | $2.0\times10^{-7}\ (300/T_g)^{0.5}$ | [98] |
| ii4 | $H^- + M^+ \rightarrow H + M$ M=$NH_x^+$, $N^+$ x=1–3 | $3.0\times10^{-6}$ | [65] |
| ii5 | $H^- + NH_4^+ \rightarrow NH_3 + H_2$ | $3.0\times10^{-6}$ | [65] |
| ii6 | $NH_2^- + NH_4^+ \rightarrow NH_3 + NH_2 + H$ | $2.0\times10^{-7}$ | [65] |
| ii7 | $NH_2^- + M^+ \rightarrow NH_2 + M$ M=$NH_x^+$, $N^+$ x=1–3 | $2.0\times10^{-7}$ | [65] |
| ii8 | $NH_2^- + H_3^+ \rightarrow NH_3 + H_2$ | $1.0\times10^{-7}$ | [65] |
| ii9 | $NH_2^- + H_3^+ \rightarrow NH_2 + H_2 + H$ | $1.0\times10^{-7}$ | [65] |

| Spontaneous emission | | Frequency | |
|---|---|---|---|
| em1 | $N_2(B) \rightarrow N_2(A) + h\nu$ | $2.0\times10^{5}$ | [60] |
| em2 | $N_2(C) \rightarrow N_2(B) + h\nu$ | $2.74\times10^{7}$ | [60] |
| em3 | $N_2(a) \rightarrow N_2(X,0) + h\nu$ | $1.8\times10^{4}$ | [60, 100] |
| em4 | $N_2(a) \rightarrow N_2(a') + h\nu$ | $1.91\times10^{2}$ | [60] |
| em5 | $N_2(w) \rightarrow N_2(a) + h\nu$ | $6.5\times10^{2}$ | [60] |

| Surface processes | | | |
|---|---|---|---|
| Sticking | | Coefficient | |
| w1 | $H + F \rightarrow H(s)$ | 1.0 | [47] |
| w2 | $N + F \rightarrow N(s)$ | 1.0 | [47] |
| w3 | $NH + F \rightarrow NH(s)$ | 1.0 | [47] |
| w4 | $NH_2 + F \rightarrow NH_2(s)$ | 1.0 | [47] |
| Dissociative adsorption | | Coefficient | |
| w5 | $N_2 + 2F \rightarrow 2N(s)$ | $3.5\times10^{-7}$ | [58] |
| w6 | $H_2 + 2F \rightarrow 2H(s)$ | $1.0\times10^{-3}$ | [58] |
| Eley-Rideal | | Coefficient | |
| w7 | $H + H(s) \rightarrow H_2 + F$ | $3.5\times10^{-3}$ | [47] |
| w8 | $N + N(s) \rightarrow N_2 + F$ | $1.2\times10^{-2}$ | [47] |
| w9 | $N + H(s) \rightarrow NH(s)$ | $1.5\times10^{-3}$ | *a* |
| w10 | $H + N(s) \rightarrow NH(s)$ | $1.2\times10^{-3}$ | *a* |
| w11 | $H + NH(s) \rightarrow NH_2(s)$ | $1.2\times10^{-3}$ | *a* |
| w12 | $NH + H(s) \rightarrow NH_2(s)$ | $1.5\times10^{-3}$ | *a* |
| w13 | $H + NH_2(s) \rightarrow NH_3 + F$ | $1.2\times10^{-3}$ | *a* |
| w14 | $H(s) + NH_2 \rightarrow NH_3 + F$ | $1.5\times10^{-3}$ | *a* |
| w15 | $H_2 + NH(s) \rightarrow NH_3 + F$ | $1.2\times10^{-4}$ | *a* |
| Langmuir-Hinshelwood | | $E_{LH}$ | |
| w16 | $NH(s) + H(s) \rightarrow NH_2(s) + F$ | 0.70 | *a* |
| w17 | $H(s) + NH_2(s) \rightarrow NH_3 + 2F$ | 0.60 | *a* |
| w18 | $N(s) + H(s) \rightarrow NH(s)$ | 1.5 | *a* |
| Surface deexcitation and neutralization | | Coefficient | |
| w19 | $N_2(A) + wall \rightarrow N_2(X,0)$ | 1.0 | [60] |
| w20 | $N_2(a') + wall \rightarrow N_2(X,0)$ | 1.0 | [60] |
| w21 | $N_2(a) + wall \rightarrow N_2(X,0)$ | 1.0 | [60] |
| w22 | $N_2(w) + wall \rightarrow N_2(X,0)$ | 1.0 | [60] |
| w23 | $N(D) + wall \rightarrow N(S)$ | 1.0 | [60] |
| w24 | $N(P) + wall \rightarrow N(S)$ | 1.0 | [60] |
| w25 | $N^+ + wall \rightarrow N$ | 1.0 | |
| w26 | $N_2^+ + wall \rightarrow N_2$ | 1.0 | |
| w27 | $N_3^+ + wall \rightarrow N_2 + N$ | 1.0 | |
| w28 | $N_4^+ + wall \rightarrow 2N_2$ | 1.0 | |
| w29 | $H^+ + wall \rightarrow H$ | 1.0 | |
| w30 | $H_2^+ + wall \rightarrow H_2$ | 1.0 | |
| w31 | $H_3^+ + wall \rightarrow H_2 + H$ | 1.0 | |
| w32 | $N_2H^+ + wall \rightarrow H + N_2$ | 1.0 | |
| w33 | $NH^+ + wall \rightarrow NH$ | 1.0 | |
| w34 | $NH_2^+ + wall \rightarrow NH_2$ | 1.0 | |
| w35 | $NH_3^+ + wall \rightarrow NH_3$ | 1.0 | |
| w36 | $NH_4^+ + wall \rightarrow NH_3 + H$ | 1.0 | |

*a*: assumed in this work (see text)
F stands for a free surface site